\documentclass[10pt]{article}

\usepackage{amssymb}
\usepackage{graphicx}
\usepackage{amsmath,amssymb}
\usepackage{color}
\usepackage{xcolor}
\usepackage{srcltx}
\usepackage{greekbf}
\usepackage{nicefrac}
\usepackage{empheq}
\usepackage{enumerate}

\usepackage{mathrsfs}

\usepackage[english]{babel}

\DeclareMathAlphabet{\mathbf}{OT1}{cmr}{bx}{it}
\DeclareMathAlphabet{\mathssb}{OT1}{cmss}{bx}{n}
\DeclareMathAlphabet{\mathssn}{OT1}{cmss}{m}{n}
\DeclareMathAlphabet{\mathub}{OT1}{cmr}{b}{n}

\DeclareMathAlphabet{\mathpzc}{OT1}{pzc}%
                                 {m}{it}

\newcommand{\<}[1]{\scriptstyle<#1>}

\newcommand\remarkname{Remark} %
\newcounter {remarkn}[section]%
\renewcommand \theremarkn {\arabic{section}.\arabic{remarkn}}%
\newenvironment{remark}{
\noindent{\emph{\remarkname}.} \rm%
}

\newcommand\remarksname{Remark} %
\newcounter {remarksn}[section]%
\renewcommand \theremarksn {\thesection.\arabic{remarksn}}%
\begin{document}

\begin{center}

\end{center}
%
%
%
%
\newcommand{\bydef}{\,\raise.050ex\hbox{\rm:}\kern-.025em\hbox{\rm=}\,}
\newcommand{\defby}{=\raise.075ex\hbox{\kern-.325em\hbox{\rm:}}\,}
\newcommand{\mtrp}  {{-\!\top}} 
\newcommand{\bdot}  {{\scriptscriptstyle\bullet}}
\def\qed{\relax\ifmmode\hskip2em \Box\else\unskip\nobreak\hskip1em $\Box$\fi}
%
%
\newcommand {\eps} {\varepsilon} 
\newcommand {\vp} {\varphi}      
\newcommand {\0} {\textbf{0}}    
\newcommand {\1} {\textbf{1}}    
%
\newcommand {\Ac}  {\mathcal{A}}
\newcommand {\Bc}  {\mathcal{B}}
\newcommand {\Cc}  {\mathcal{C}}
\newcommand {\Dc}  {\mathcal{D}}
\newcommand {\Ec}  {\mathcal{E}}
\newcommand {\Fc}  {\mathcal{F}}
\newcommand {\Gc}  {\mathcal{G}}
\newcommand {\Hc}  {\mathcal{H}}
\newcommand {\Kc}  {\mathcal{K}}
\newcommand {\Ic}  {\mathcal{I}}
\newcommand {\Jc}  {\mathcal{J}}
\newcommand {\Lc}  {\mathcal{L}}
\newcommand {\Mc}  {\mathcal{M}}
\newcommand {\Nc}  {\mathcal{N}}
\newcommand {\Oc}  {\mathcal{O}}
\newcommand {\Pc}  {\mathcal{P}}
\newcommand {\Rc}  {\mathcal{R}}
\newcommand {\Sc}  {\mathcal{S}}
\newcommand {\Tc}  {\mathcal{T}}
\newcommand {\Uc}  {\mathcal{U}}
\newcommand {\Vc}  {\mathcal{V}}
\newcommand {\Wc}  {\mathcal{W}}
\newcommand {\Zc}  {\mathcal{Z}}
%
%
\newcommand {\ab} {\mathbf{a}}
\newcommand {\bb} {\mathbf{b}}
\newcommand {\cb} {\mathbf{c}}
\newcommand {\db} {\mathbf{d}}
\newcommand {\eb} {\mathbf{e}}
\newcommand {\fb} {\mathbf{f}}
\newcommand {\gb} {\mathbf{g}}
\newcommand {\hb} {\mathbf{h}}
\newcommand {\ib} {\mathbf{i}}
\newcommand {\kb} {\mathbf{k}}
\newcommand {\lb} {\mathbf{l}}
\newcommand {\mb} {\mathbf{m}}
\newcommand {\nb} {\mathbf{n}}
\newcommand {\pb} {\mathbf{p}}
\newcommand {\qb} {\mathbf{q}}
\newcommand {\rb} {\mathbf{r}}
\renewcommand {\sb} {\mathbf{s}}
\newcommand {\tb} {\mathbf{t}}
\newcommand {\xb} {\mathbf{x}}
\newcommand {\ub} {\mathbf{u}}
\newcommand {\vb} {\mathbf{v}}
\newcommand {\wb} {\mathbf{w}}
\newcommand {\zb} {\mathbf{z}}
\newcommand {\Ab} {\mathbf{A}}
\newcommand {\Bb} {\mathbf{B}}
\newcommand {\Cb} {\mathbf{C}}
\newcommand {\Db} {\mathbf{D}}
\newcommand {\Eb} {\mathbf{E}}
\newcommand {\Fb} {\mathbf{F}}
\newcommand {\Gb} {\mathbf{G}}
\newcommand {\Hb} {\mathbf{H}}
\newcommand {\Kb} {\mathbf{K}}
\newcommand {\Jb} {\mathbf{J}}
\newcommand {\Ib} {\mathbf{I}}
\newcommand {\Lb} {\mathbf{L}}
\newcommand {\Mb} {\mathbf{M}}
\newcommand {\Nb} {\mathbf{N}}
\newcommand {\Ob} {\mathbf{O}}
\newcommand {\Pb} {\mathbf{P}}
\newcommand {\Qb} {\mathbf{Q}}
\newcommand {\Rb} {\mathbf{R}}
\newcommand {\Sb} {\mathbf{S}}
\newcommand {\Tb} {\mathbf{T}}
\newcommand {\Vb} {\mathbf{V}}
\newcommand {\Wb} {\mathbf{W}}
\newcommand {\Xb} {\mathbf{X}}
\newcommand {\Zb} {\mathbf{Z}}
%

%
%
\newcommand {\ax} {\mathrm{a}}
\newcommand {\bx} {\mathrm{b}}
\newcommand {\cx} {\mathrm{c}}
\newcommand {\dx} {\mathrm{d}}
\newcommand {\ex} {\mathrm{e}}
\newcommand {\fx} {\mathrm{f}}
\newcommand {\gx} {\mathrm{g}}
\newcommand {\hx} {\mathrm{h}}
\newcommand {\kx} {\mathrm{k}}
\newcommand {\lx} {\mathrm{l}}
\newcommand {\mx} {\mathrm{m}}
\newcommand {\nx} {\mathrm{n}}
\newcommand {\ox} {\mathrm{o}}
\newcommand {\px} {\mathrm{p}}
\newcommand {\qx} {\mathrm{q}}
\newcommand {\rx} {\mathrm{r}}
\newcommand {\sx} {\mathrm{s}}
\newcommand {\tx} {\mathrm{t}}
\newcommand {\xx} {\mathrm{x}}
\newcommand {\yx} {\mathrm{y}}
\newcommand {\wx} {\mathrm{w}}
\newcommand {\ux} {\mathrm{u}}
\newcommand {\vx} {\mathrm{v}}
\newcommand {\zx} {\mathrm{z}}
\newcommand {\Ax} {\mathrm{A}}
\newcommand {\Bx} {\mathrm{B}}
\newcommand {\Cx} {\mathrm{C}}
\newcommand {\Dx} {\mathrm{D}}
\newcommand {\Ex} {\mathrm{E}}
\newcommand {\Fx} {\mathrm{F}}
\newcommand {\Gx} {\mathrm{G}}
\newcommand {\Hx} {\mathrm{H}}
\newcommand {\Kx} {\mathrm{K}}
\newcommand {\Jx} {\mathrm{J}}
\newcommand {\Ix} {\mathrm{I}}
\newcommand {\Lx} {\mathrm{L}}
\newcommand {\Mx} {\mathrm{M}}
\newcommand {\Nx} {\mathrm{N}}
\newcommand {\Ox} {\mathrm{O}}
\newcommand {\Px} {\mathrm{P}}
\newcommand {\Qx} {\mathrm{Q}}
\newcommand {\Rx} {\mathrm{R}}
\newcommand {\Sx} {\mathrm{S}}
\newcommand {\Tx} {\mathrm{T}}
\newcommand {\Vx} {\mathrm{V}}
\newcommand {\Wx} {\mathrm{W}}
\newcommand {\Xx} {\mathrm{X}}
\newcommand {\Yx} {\mathrm{Y}}
\newcommand {\Zx} {\mathrm{Z}}
%

%
%
\newcommand {\Real} {\mathbb{R}}
\newcommand {\Aos} {\mbox{$\scriptstyle\mathbb{A}$}}
\newcommand {\Ao} {\mathbb{A}}
\newcommand {\Bo} {\mathbb{B}}
\newcommand {\Co} {\mathbb{C}}
\newcommand {\Cop} {\mbox{$\scriptstyle\mathbb{C}$}}
\newcommand {\Do} {\mathbb{D}}
\newcommand {\Fo} {\mathbb{F}}
\newcommand {\Go} {\mathbb{G}}
\newcommand {\Io} {\mathbb{I}}
\newcommand {\Mo} {\mathbb{M}}
\newcommand {\Ko} {\mathbb{K}}
\newcommand {\No} {\mathbb{N}}
\newcommand {\Po} {\mathbb{P}}
\newcommand {\Qo} {\mathbb{Q}}
\newcommand {\Ro} {\mathbb{R}}
\newcommand {\So} {\mathbb{S}}
\newcommand {\To} {\mathbb{T}}
\newcommand {\Vo} {\mathbb{V}}
\newcommand {\Zo} {\mathbb{Z}}
%
\newcommand {\Ds} {\mathscr{D}}
\newcommand {\Ms} {\mathscr{M}}
\newcommand {\Ns} {\mathscr{N}}
\newcommand {\Vs} {\mathscr{V}}
\newcommand {\Xes} {\mathscr{X}}
\newcommand {\ges} {\mathscr{g}}
\newcommand {\wes} {\mathscr{W}}
%
%
%
\newcommand {\aef} {\mathfrak{a}}
\newcommand {\fef} {\mathfrak{f}}
\newcommand {\gef} {\mathfrak{g}}
\newcommand {\hef} {\mathfrak{h}}
\newcommand {\mef} {\mathfrak{m}}
\newcommand {\nef} {\mathfrak{n}}
\newcommand {\kef} {\mathfrak{k}}
\newcommand {\wef} {\mathfrak{w}}
\newcommand {\sef} {\mathfrak{s}}
\newcommand {\zef} {\mathfrak{z}}
\newcommand {\Def} {\mathfrak{D}}
\newcommand {\Fef} {\mathfrak{F}}
\newcommand {\Mef} {\mathfrak{M}}
\newcommand {\Nef} {\mathfrak{N}}
\newcommand {\Ref} {\mathfrak{R}}
\newcommand {\Sef} {\mathfrak{S}}
\newcommand {\Xef} {\mathfrak{X}}
%
%
\newcommand {\att} {\mathtt{a}}
\newcommand {\btt} {\mathtt{b}}
\newcommand {\ctt} {\mathtt{c}}
\newcommand {\dtt} {\mathtt{d}}
\newcommand {\ftt} {\mathtt{f}}
\newcommand {\gtt} {\mathtt{g}}
\newcommand {\mtt} {\mathtt{m}}
\newcommand {\ntt} {\mathtt{n}}
\newcommand {\htt} {\mathtt{h}}
\newcommand {\ptt} {\mathtt{p}}
\newcommand {\qtt} {\mathtt{q}}
\newcommand {\rtt} {\mathtt{r}}
\newcommand {\stt} {\mathtt{s}}
\newcommand {\ttt} {\mathtt{t}}
\newcommand {\vtt} {\mathtt{v}}
\newcommand {\wtt} {\mathtt{w}}
\newcommand {\ztt} {\mathtt{z}}
\newcommand {\Ftt} {\mathtt{F}}
\newcommand {\Stt} {\mathtt{P}}
\newcommand {\Wtt} {\mathtt{W}}
%
%
%
%
\newcommand {\alfab}     {\mathbf{\alpha}}
\newcommand {\betab}     {\mathbf{\beta}}
\newcommand {\gammab}    {\mathbf{\gamma}}
\newcommand {\deltab}    {\mathbf{\delta}}
\newcommand {\epsilonb}  {\mathbf{\epsilon}}
\newcommand {\epsb}      {\mathbf{\varepsilon}}
\newcommand {\zetab}     {\mathbf{\zeta}}
\newcommand {\etab}      {\mathbf{\eta}}
\newcommand {\tetab}     {\mathbf{\teta}}
\newcommand {\vtetab}    {\mathbf{\vartheta}}
\newcommand {\iotab}     {\mathbf{\iota}}
\newcommand {\kappab}    {\mathbf{\kappa}}
\newcommand {\lambdab}   {\mathbf{\lambda}}
\newcommand {\mub}       {\mathbf{\mu}}
\font\mbo=cmmib10 scaled \magstephalf
\newcommand{\nub}   {\hbox{\mbo {\char 23}}}
\newcommand {\csib}      {\mathbf{\xi}}
\newcommand {\xib}      {\mathbf{\xi}}
\newcommand {\pib}       {\mathbf{\pi}}
\newcommand {\varrhob}   {\mathbf{\varrho}}
\newcommand {\sigmab}    {\hbox{\mbo {\char 27}}}
\newcommand{\taub}   {\hbox{\mbo {\char 28}}}
\newcommand {\upsilonb}  {\mathbf{\upsilon}}
\newcommand {\phib}      {\mathbf{\phi}}
\newcommand {\varphib}   {\mathbf{\varphi}}
\newcommand {\chib}      {\mathbf{\chi}}
\newcommand {\psib}       {\mathbf{\psi}}
\newcommand {\omegab}    {\mathbf{\omega}}
%
%
%
%
\newcommand {\Gammab}    {\mathbf{\Gamma}}
\newcommand {\Deltab}    {\mathbf{\Delta}}
\newcommand {\Tetab}     {\mathbf{\Theta}}
\newcommand {\Lambdab}   {\mathbf{\Lambda}}
\newcommand {\Csib}      {\mathbf{\Xi}}
\newcommand {\Pib}       {\mathbf{\Pi}}
\newcommand {\Sigmab}    {\mathbf{\Sigma}}
\newcommand {\Phib}      {\mathbf{\Phi}}
\newcommand {\Psib}      {\mathbf{\Psi}}
\newcommand {\Omegab}    {\mathbf{\Omega}}
%
%
\newcommand {\Lin} {\mathbb{L}\mathtt{in}}
\newcommand {\Sym} {\mathbb{S}\mathtt{ym}}
\newcommand {\Psym} {\mathbb{PS}\mathtt{ym}}
\newcommand {\Skw} {\mathbb{S}\mathtt{kw}}
\newcommand {\SO} {\mathcal{SO}}
\newcommand {\GL} {\mathcal{G}l}
\newcommand {\Rot} {\mathbb{R}\mathtt{ot}}
%
%
\newcommand {\tr}[1]{\mbox{tr}\, #1}
\newcommand {\psym} {\mbox{sym}}
\newcommand {\pskw} {\mbox{skw}}
\newcommand {\win}[2] {( #1 \cdot #2 )_\wedge }
\newcommand {\modulo}[1] {\left|#1\right|}
\newcommand {\sph} {\mbox{sph}}
\newcommand {\dev} {\mbox{dev}}
\newcommand {\sgn} {\mbox{sgn}}
\newcommand {\lin} {\mbox{Lin}}
%
%
\newcommand{\dvg} {\mathrm{div}\,}
\newcommand{\dvgt} {\mathrm{d{\widetilde i}v}\,}    
\newcommand{\grd} {\mathrm{grad}\,}    
\newcommand{\Grd} {\mathrm{Grad}\,}    
\newcommand{\dl}  {\delta}             
\def\gradtwo{\mathord{\nabla^{\scriptscriptstyle(2)}}}
\def\Div{\mathop{\hbox{Div}}}
\def\div{\mathop{\hbox{div}}}
\def\mis{\mathop{\hbox{mis}}}
\def\eps{\varepsilon}
%
%

\newcommand\ph{\varphi}
\newcommand{\adj} {\att\dtt}     

\newcommand{\va}{\mathbf{a}}
\newcommand{\vn}{\mathbf{\nu}}
\newcommand{\vt}{\mathbf{\tau}}
\newcommand{\dn}{\partial_{\mathbf{\nu}}}
\newcommand{\dt}{\partial_{\mathbf{\tau}}}
\newcommand{\ord}{\scriptscriptstyle}
\newcommand{\tD}{\mathbf{E}}  
\newcommand{\tS}{\mathbf{S}}  
\newcommand{\tE}{\mathbb{C}}
\newcommand{\tPf}{\mathbb{P}}
\newcommand{\tPr}{\mathbf{P}}
\newcommand{\tC}{\mathbf{C}}
\newcommand{\tP}{{\scriptstyle\mathbb{C}}}   
\newcommand{\tDl}{\mathbf{C}}      

\newcommand{\Cuno}{\mathbf{c}_1}
\newcommand{\Cdue}{\mathbf{c}_2}
\newcommand{\veralf}{\mathbf{c}_\alpha} 
\newcommand{\verbet}{\mathbf{c}_\beta} 
\newcommand{\vz}{\mathbf{z}}
\newcommand{\sym}{\mathop{\mathrm{sym}}}

\newcommand{\f}{f}
\newcommand{\g}{g}
\newcommand{\h}{h}
\newcommand{\w}{w}
\renewcommand{\l}{l}

\renewcommand{\r}{r}
\newcommand{\s}{s}

\newcommand{\autof}{\mathrm{\skew 0\overline{w}}}
\newcommand{\W}{W}

\newcommand{\df}{f'}
\newcommand{\dg}{g'}
\renewcommand{\dh}{h'}
\newcommand{\ddf}{f''}
\newcommand{\ddg}{g''}
\newcommand{\ddh}{h''}

\newcommand\modv[1]{|{#1}|}

\newcommand\arr[1]{\overrightarrow{#1}}
\newcommand{\cartref}{\{O;x_1,x_2,x_3\}}
\newcommand{\orthframe}{({\bf e}_1, {\bf e}_2, {\bf e}_3)}


\newcommand\email[1]{\texttt{#1}}
\newcommand\at{:}

\begin{center}
 {\bf \Large
On a Nanoscopically-Informed Shell Theory of Single-Wall Carbon
Nanotubes}
\end{center}
\medskip

\begin{center}
{\large Chandrajit Bajaj$^{(1)}$ \; Antonino Favata$^{(2)}$ \;
                Paolo Podio-Guidugli$^{(2)}$
}\end{center}

\begin{center}
 \noindent $^{(1)}$Center for Computational Visualization,
 Institute for Computational Engineering and Sciences The
 University of Texas at Austin\footnote {201 East 24th Street, Austin, TX
 78712, USA \\
{\null} \quad \ Email:
\begin{minipage}[t]{30em}
\email{bajaj@ices.utexas.edu}
\end{minipage}}
\small
\end{center}

\begin{center}
 \noindent $^{(2)}$Dipartimento di Ingegneria Civile, Universit\`a di Roma Tor Vergata\footnote{Via Politecnico 1, 00133 Rome, Italy. \\
{\null} \quad \ Email:
\begin{minipage}[t]{30em}
\email{favata@ing.uniroma2.it} (A. Favata)\\
\email{ppg@uniroma2.it} (P. Podio-Guidugli)
\end{minipage}}
\small
\end{center}
\medskip

\begin{abstract}
\noindent {\footnotesize This
paper proposes a bottom-up sequence of modeling steps leading to a
nanoscopically informed continuum, and as such macroscopic, theory
of single-walled carbon nanotubes (SWCNTs). We provide a
description of the geometry of the two most representative types
of SWCNTs, armchair (A-) and zigzag (Z-), of their modules and of
their elementary bond units. We believe ours to be the simplest shell
theory that accounts accurately for the linearly elastic response of both
A- and Z- CNTs. In fact, our theory can be shown to fit
SWCNTs of whatever chirality; its main novel feature is perhaps
the proposition of chirality-dependent concepts of effective
thickness and effective radius.
\medskip

}
\end{abstract}

\tableofcontents
\section{Introduction} Carbon has  a large variety of
allotropic forms, natural and artificial: graphite, diamond, fullerene, carbon fibers,
carbon nanotubes (CNTs), and others. The relative C-C bond complexes, diverse as they are by spatial dimensions and geometry, correspond to diverse hybrid electron states, made possible by combination of the electrons occupying the most external shells in a C atom;  the nature and the strength of its C-C bond complex decide the mechanical properties of a given allotrope at any scale.

This paper proposes a bottom-up sequence of modeling steps leading to a nanoscopically informed continuum, and as such macroscopic, theory of single-walled carbon nanotubes (SWCNTs).
Recently, modeling procedures of the same type have been adopted by various
authors \cite{Li,CG,SL,Od,Od1,GC,WD}, with application to CNTs and
polymer composites in mind. Related multiscale modeling papers, such as \cite{BLBL} and \cite{LBL}, cover a variety of tools for studying both quantum and classical models of atomic systems in crystalline phase, and propose techniques to connect the microscopic models to the continuum limits. They suggest that, for non-periodic systems, the key challenge lies in carefully defining average energy at the microscopic level and connecting it to a macroscopic energy. Another related paper is \cite{MB}, that studies the mechanics of materials with periodic micro-structures; here, lattice structures are divided into two classes: truss materials modeled as beams with axial and bending stiffness, and frame materials modeled with pin-joints at the nodes.

There are of course differences in concept
ingredients and developments in the cited papers, but their goal is one and the same: \emph{to deduce the targeted macroscopic mechanical behavior from the available nanoscopic chemical-physical
information.} In our present paper, a bridge between the nanoscopic and the macroscopic viewpoints is created by means of an intermediate passage, that we term \emph{meta-nanoscopic}, in a fashion that we now describe.
At the nanoscopic scale, the
bonding and non-bonding energies keeping CNTs together are evaluated in the fashion of molecular
mechanics (MM) (Section \ref{N}). In the following section, a
mechanical caricature of a SWCNT as an orderly arrangement of
 pin-jointed sticks and (axial and spiral) springs is
drawn, within the framework of discrete structure mechanics (DSM). This requires a preliminary careful description of the geometry of the two most representative types of SWCNTs, armchair (A-) and zigzag (Z-), of their modules and of their elementary bond units; in such description, the nanoscopic parameters are the chirality index and the length of the C-C bond. The viewpoints
of MM and DSM are connected by equating the
energies per bond, for each of which an approximate quadratic expression is accepted: two types of bonding energies -- hence two purely constitutive parameters, the stiffness constants of the two types of springs --   enter the expression for the DSM energy of a bond unit; they are taken equal to the two corresponding nanoscopic stiffness parameters, regarded as a known input from MM. This simple measure opens the way to the posing and solving, within the framework of DSM, of a number of equilibrium problems for SWCNTs (Section \ref{bvps}).

The step from discrete to continuous structure mechanics (CSM) is harder. To take it, a DSM model of SWCNT has to be assimilated to a suitable model taken from the library of CSM.
The main difficulty in doing so resides in the number and nature of the parameters needed to specify a CSM model: no matter what rod or shell theory one chooses, the relative stiffness notion consists of a list of more than two combinations of  both \emph{constitutive} and \emph{geometric} parameters, the latter reflecting theory-specific concepts of \emph{thinness} and \emph{slenderness}. Now, while there is a natural nanoscopic notion of slenderness for SWCNTs, to assess their thinness is a controversial issue, because a preliminary estimate of their \emph{effective thickness} is required (these matters are dealt with in  Section \ref{slen}); and, no doubt both slenderness and thinness of a SWCNT depend on its chirality.
Therefore, one is faced with the problem of determining the chirality-dependent parameters of the continuum theory of his choice in terms of only two nanoscopic constitutive parameters and in the absence of an unequivocal  nanoscopic concept of thinness.

For reasons that are discussed at length in \cite{Fa2} and recapitulated in Section \ref{macro}, the CSM theory we adopt is \emph{a linearly elastic theory of  orthotropic shells}, where the independent constitutive parameters are four and the geometric parameters two. To determine all those parameters, firstly we solve the system of equilibrium equations ruling the \emph{axial-traction} and \emph{radial-loading} DSM problems for both A- and Z- CNTs, and the \emph{torsion} problem  for ACNTs (Sections \ref{ata}--\ref{325}); interestingly, the solutions we find hold for whatever slenderness. Secondly, we import from \cite{Fa2} the analytic solutions there found for the same equilibrium problems, when set within the framework of the above mentioned shell theory. While the torsion solution is insensitive to slenderness, the axial-traction and radial-loading solutions are; for simplicity, we confine attention to the case of slender shells. Thirdly, for each of these boundary-value problems, both in the discrete and in the continuous formulation, we regard the involved nanotubes as cylindrical \emph{probes}, whose relevant stiffnesses and Poisson-like contraction ratios we define in the fashion of standard rod theory (Sections \ref{61} and \ref{62}, respectively). Finally, in Section \ref{costpar}, the sought-for parameters are found by equating the continuous and discrete stiffnesses and contraction ratios; since we have an unusually rich crop of solutions at our disposal, this can be done in more than one way, with convincingly consistent results.

We believe ours to be the simplest shell theory that accounts accurately for the elastic response of both A- and Z- CNTs. In fact, modulo a modest generalization to be presented elsewhere, our theory can be shown to fit SWCNTs of whatever chirality; its main novel feature is perhaps the proposition of chirality-dependent concepts of effective thickness and effective radius; for a discussion of this issue, and others, we refer the reader to our conclusion section.

\section{Nanoscale Energies}\label{N}
At nanometer scale, atomic
interactions are modeled by either quantum mechanics (QM) or
molecular mechanics (MM), in both cases by striving to capture how and to what extent the system's energy varies with changes in atomic positions. QM
can describe rigorously the electronic structure of a material
complex, but its computational cost quickly becomes prohibitive
as the number of atoms involved increases. MM is based on the \emph{Born--Oppenheimer approximation} for the Hamiltonian of a collection of heavy and light particles, a cornerstone of computer simulations.
Consistent with this simplifying assumption, the total energy $U$ of such a material complex
is expressed as a sum of two terms:
$$
U=U_b+U_{nb},
$$
where $U_{b}$ and $U_{nb}$, denote, respectively, the
\emph{bonding energies} and the \emph{non-bonding energies}. The latter account  for van der Waals and Coulomb interactions. However, when modeling SWCNT, the contribution of Coulomb interactions is usually neglected since interactions are predominantly between essentially neutral carbon atoms. Within the same modeling context, the former consist of four parts:
$$
U_b=U_{\rho}+U_{\theta}+U_\omega+U_\tau,
$$
where $U_\rho, U_\theta, U_\omega$ and  $U_\tau$ denote the
energies respectively associated with \emph{stretching} (of a
covalent bond), \emph{angle variation} (between two covalent
bonds), \emph{torsion} (around bonds), and the so-called \emph{improper torsion} (see e.g. \cite{Od}).
The third and fourth contributions are considered
negligible when compared with $U_\rho$ and $U_\theta$, whose harmonic approximations are:
$$
\begin{aligned}
&U_\rho=\frac{1}{2}\sum k_\rho(\rho-\rho^{\rm ref})^2,\\
&U_\theta=\frac{1}{2}\sum k_\theta(\theta-\theta^{\rm
ref})^2.
\end{aligned}
$$
Thus, the
only bond-stiffness constants of importance are  $k_\rho$ and
$k_\theta$; they can be obtained by \emph{ab initio} QM
evaluations or fitted to experiments. The values we shall employ in our numerical computations,
$$
k_\rho=742\;\, \textrm{nN}/\textrm{nm}\quad\textrm{and}\quad k_\theta=1.142\;\, \textrm{nN}\times\textrm{nm},
$$
are the same as in \cite{CG} and  \cite{SL}.

\section{The Geometry of SWCNTs, from A to Z}\label{nanosc}
\begin{figure}[h]
\centering
\includegraphics[scale=0.5]{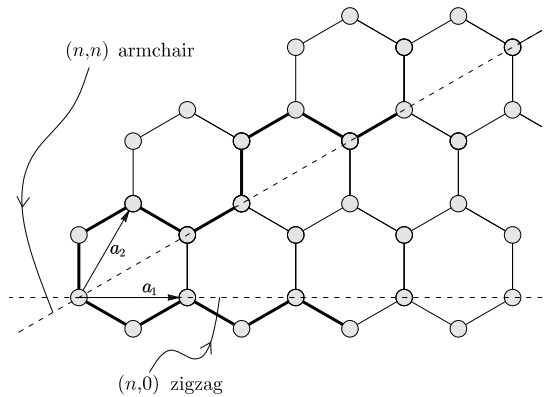}
\caption{Graphene and its zigzag and armchair chiral axes.}
\label{graph}
\end{figure}

\begin{figure}[h]
\centering
\includegraphics[scale=0.75]{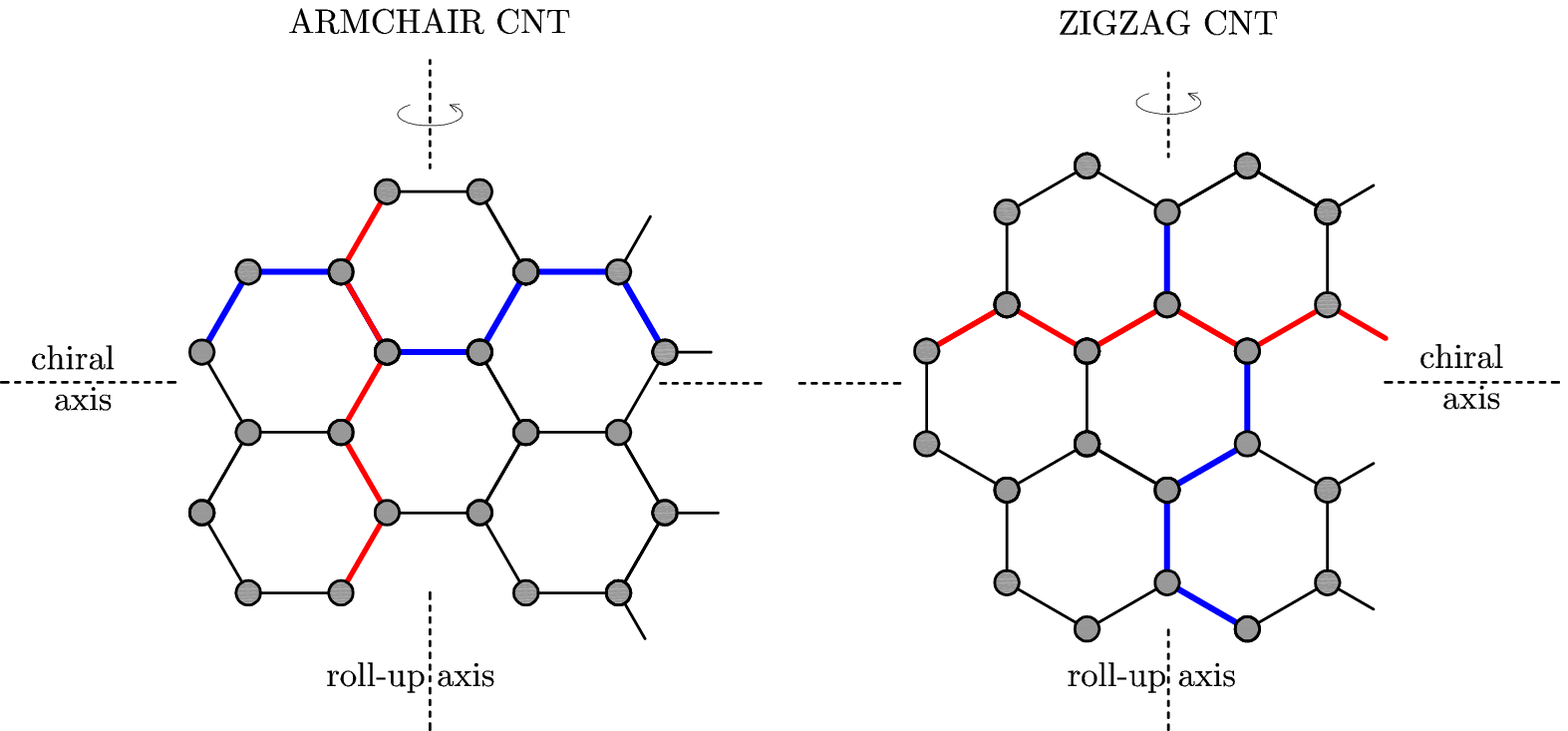}
\caption{Graphenes to be rolled up into armchair (left) and zigzag (right) single-wall CNTs. Note the orthogonality of zigzag (red) and armchair (blue) atom sequences.} \label{ort}
\end{figure}

CNTs were discovered in 1991 by S. Iijima
\cite{Ii}. They look like right cylinders with approximately emispherical caps due to the actual manifacturing procedures, and can be single- or multi-wall. In imagination, SWCNTs can
be obtained by rolling up into a cylindrical shape a
\emph{graphene}, that is, a monolayer flat sheet of graphite that can be visualized as a two-dimensional lattice with hexagonal unit cell.
There are many ways to roll a graphene up, sorted by introducing a geometrical
object, the \emph{chiral vector}:
$$
\chib=n\ab_1+m\ab_2
$$
(Fig. \ref{graph}).
Having fixed $\ab_1$ and $\ab_2$, the pairs of integers $(n,m)$
specifies the chirality of a CNT; $(n,0)-$ and
$(n,n)-$nanotubes are called, respectively, \emph{zigzag} and
\emph{armchair} (Fig. \ref{ort}; roll-up axes are chosen perpendicular to chiral axes).\footnote{The \emph{chiral angle} is defined to be $\hat\chi(n,m)=\arctan(\sqrt 3 \frac{m}{m+2n})$; thus, in particular, the chiral angle of an armchair CNT is $\pi/6$ radians.}

Chirality influences all geometrical and mechanical properties of a CNT; for example,  the radius of the cylinder on which a $(n,m)-$SWNT sits is:
\begin{equation}\label{raggio}
\rho_0=\widehat\rho_0(n,m):=\frac{1}{2\pi}\,\sqrt{3(n^2+nm+m^2)}\,s,
\end{equation}
where $s$ the side length of a graphene cell; and, chirality influences
the effective extension and shear moduli and the buckling strain
\cite{CC,C,G,Ru,ZTW}; needless to say, for multi-wall carbon nanotubes (MWCNTs), the chirality of the inner walls
is hard to inspect.

In this paper, \emph{we concentrate on SWCNTs, armchair
and zigzag}; for short, we designate them by ACNTs and ZCNTs,
respectively. The geometry of their undeformed reference
configurations is described just below. To get acquainted with
their size, it is convenient to start from a graphene cell: its
side length $s$ equals the C--C bond length,  about 0.142 nm (that
is, roughly, the diameter of a C atom); consequently, two opposite
C atoms are spaced by $(1+2\cos\pi/3)s =0.283$ nm = the cell's
diameter; and, equally placed C atoms belonging to adjacent cells
are spaced by $|\ab_1|=|\ab_2|=(2\sin\pi/3) s = 0.2456$ nm, the
distance of two parallel sides of a cell (to fix ideas, the
spacing between two adjacent cylinders in a MWNT is about 0.34
nm).

\subsection{Module and bond unit of an ACNT}
\begin{figure}[h]
\centering
\includegraphics[scale=.75]{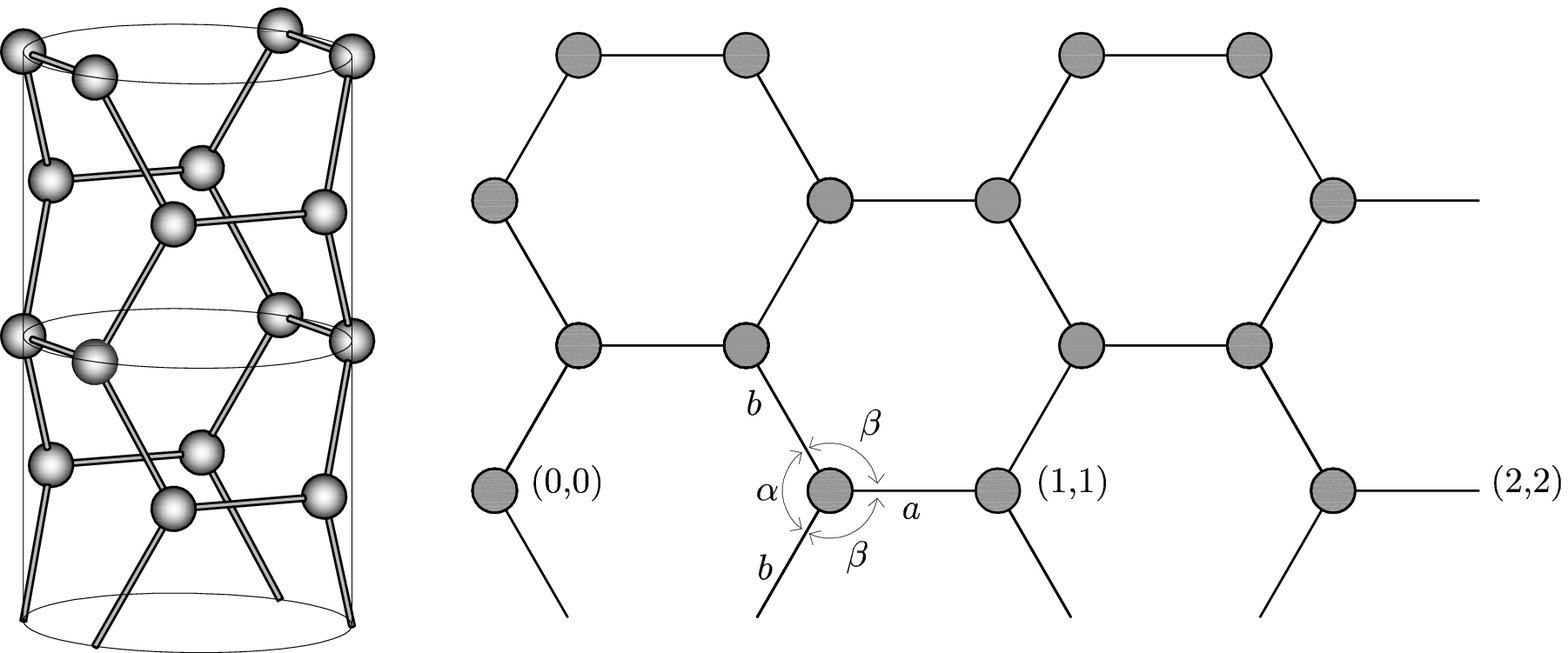}
\caption{The module of a $(2,2)-$CNT, rolled and not.} \label{3d}
\end{figure}

\begin{figure}[h]
\centering
\includegraphics[scale=.8]{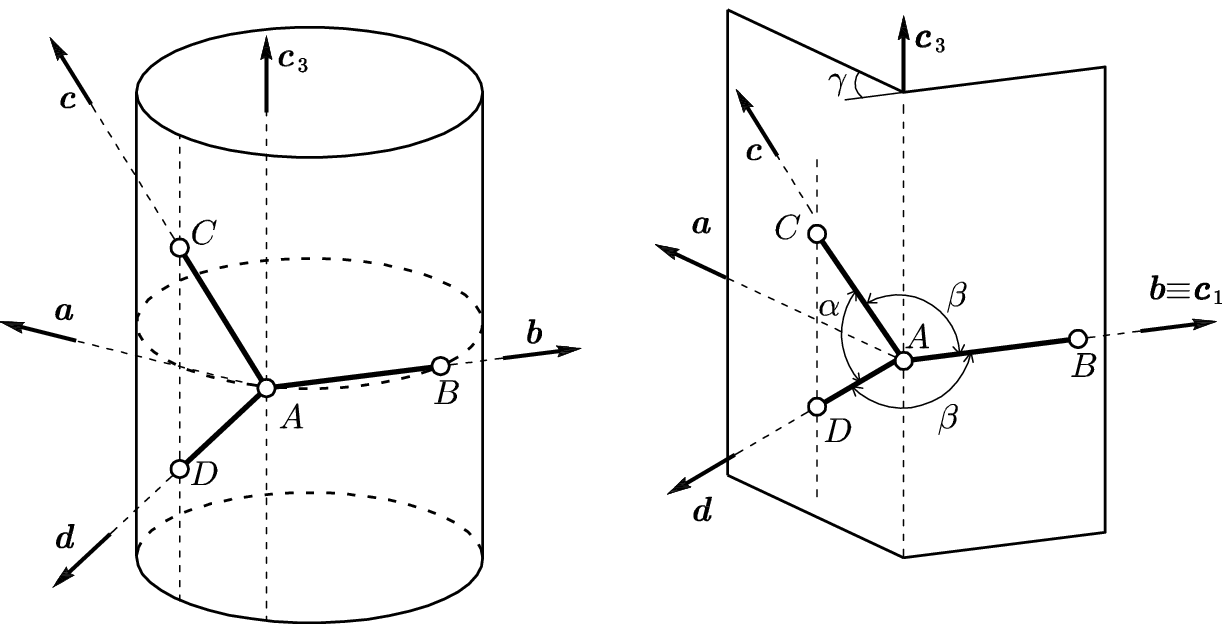}
\caption{The geometry of an ABU: front view.} \label{fram}
\end{figure}

\begin{figure}[h]
\centering
\includegraphics[scale=0.9]{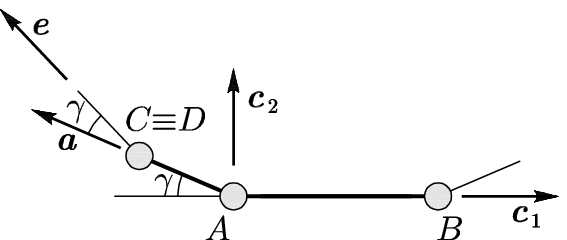}
\caption{The geometry of an ABU: top-down view.} \label{TABU}
\end{figure}

Figure \ref{3d} is an atom-and-bond cartoon of the module of the ACNT having the smallest radius  one can think of ($n=2$), in its reference configuration.
Whatever $n$, the \emph{geometrically necessary radius}, that is, the radius of the cylinder on which the centers of the C atoms are placed, is:
\begin{equation}\label{rozzo}
\widehat\rho_0(n,n)=:\rho_0^A(n)=\frac{3}{2\pi}\,n\,s
\end{equation}
(cf. \eqref{raggio}); the total number of atoms per module is:
\[
N^{AM}=4\times2n=8n.
\]
distinguish two types of sticks, $a$ horizontal and $b$ oblique,
and two types of angles between sticks, denoted by $\alpha$ and $\beta$; in a module, there are:
\begin{equation}
\begin{aligned}
N_a^{AM}&=4\times n\quad \qquad \quad \; \; \;\, \textrm{type $a$ sticks};\\
N_b^{AM}&=4\times 2n=8n \qquad \textrm{type $b$ sticks};\\
N_\alpha^{AM}&=4\times 2n=8n \quad\;\;\;\, \textrm{type $\alpha$ angles};\\
N_\beta^{AM}&=4\times 4n=16n \quad\;\; \textrm{type $\beta$ angles}.
\end{aligned}
\end{equation}
Three adjacent sticks, two of type $b$ and one of type $a$, form an armchair bond unit (ABU).
We make the geometry of an undeformed ABU precise with reference to Figures \ref{fram} and \ref{TABU},
where atom $A$ is sitting at the origin of
an orthogonal cartesian frame of unit vectors $\cb_i$, with $\cb_1\equiv\bb$. Note that
\begin{equation}\label{cos}
\cos\beta=-\cos\frac{\alpha}{2}\cos\gamma, \quad \gamma=\frac{\pi}{2n},
\end{equation}
and that the three next neighbors of $A$ sit at the ends of the relative position vectors:
\begin{equation}\label{posiz}
\begin{aligned}
&s^{-1}\pb_B:=\bb=\cb_1,\\
&s^{-1}\pb_C:=\cb=\cos\frac{\alpha}{2}\ab+\sin\frac{\alpha}{2}\cb_3,\\
&s^{-1}\pb_D:=\db=\cos\frac{\alpha}{2}\ab-\sin\frac{\alpha}{2}\cb_3,
\end{aligned}
\end{equation}
where the unit vector
\begin{equation}\label{avect}
\ab=-\cos\gamma\cb_1+\sin\gamma\cb_2
\end{equation}
is orthogonal to
$\cb_3$; moreover,
\begin{equation}\label{evect}
\eb=-\cos 2\gamma\,\cb_1+\sin 2\gamma\,\cb_2.
\end{equation}
\remark  One may wonder why we have not set $\alpha=2\pi/3$ in the formulas just above. This is in fact the case for the undeformed configuration, when we deduce from \eqref{cos} that
\begin{equation}\label{betaA}
\beta=\widehat\beta^A(n)=\pi-\arccos\left(\frac{1}{2}\cos\frac{\pi}{2n}\right).
\end{equation}
Our reason is that $\eqref{cos}_1$ holds also when the applied
loads induce deformations that, while changing angles of both
types $\alpha$ and $\beta$, do not change angle $\gamma$ and
preserve the mirror symmetry with respect to the plane through $A$
orthogonal to $\cb_3$ of the position vectors of atoms $C$ and $D$
(e.g., these conditions are met in the axial-traction problem we
are going to study in Section \ref{ata}, but they are not in the
torsion problem studied in Section \ref{torz}). We record here for
later use an expression of $\Delta\beta$ as a function of
$\Delta\alpha$ that follows by differentiation of $\eqref{cos}_1$
while keeping $\gamma$ fixed:
\begin{equation}\label{deltavari}
\Delta\beta=-\widehat\delta^A(\alpha,\beta,\gamma)\Delta\alpha, \quad
\widehat\delta^A(\alpha,\beta,\gamma):=\frac{\sin\frac{\alpha}{2}}{2\sin\beta}\,\cos\gamma=\frac{\sqrt 3}{4\sin\beta}\,\cos\gamma\,.
\end{equation}
(cf. eq. (14) of \cite{CG}).
\subsection{Module and bond unit of a ZCNT}
\begin{figure}[h]
\centering
\includegraphics[scale=0.8]{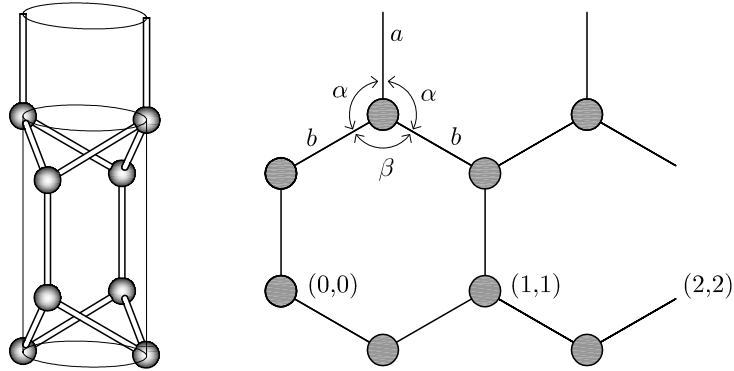}
\caption{The module of a $(2,0)-$CNT.} \label{BU_zigzag}
\end{figure}

\begin{figure}[h]
\centering
\includegraphics[scale=0.8]{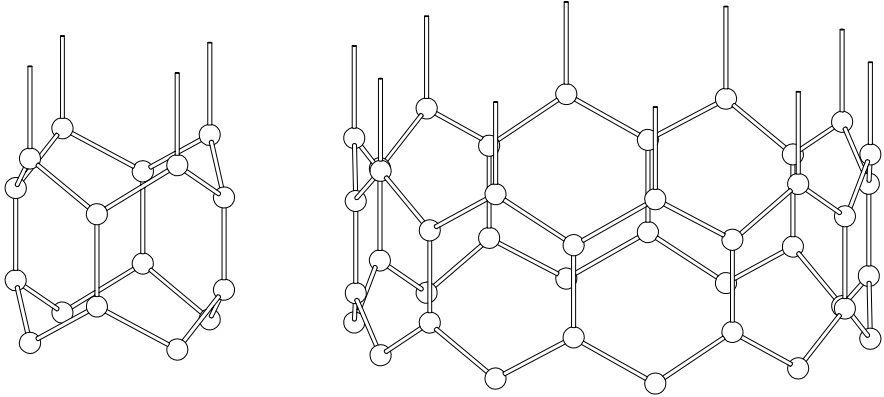}
\caption{A $(4,0)-$ and a $(10,0)-$CNT.} \label{10_zig}
\end{figure}

The module of a
ZCNT of smallest radius ($n=2$) is drawn in Figure \ref{BU_zigzag}; Figure \ref{10_zig} is relative to the cases $n=4$ and $n=10$.
\noindent The cylinder's radius and the total number of atoms per
module are, respectively,
\begin{equation}\label{roz}
\widehat\rho_0(n,0)=:\rho_0^Z(n)=\frac{\sqrt3}{2\pi}\,n\,s\quad\textrm{and}\quad
N^{ZM}=4\times n=4n.
\end{equation}
In each module, there are:
\begin{equation}
\begin{aligned}
N_a^{ZM}&=2\times n=2n\quad  \;\; \; \; \;\,\,  \textrm{type $a$ sticks};\\
N_b^{ZM}&=2\times 2n=4n \qquad \textrm{type $b$ sticks};\\
N_\alpha^{ZM}&=2\times 4n=8n \quad\;\;\;\, \textrm{type $\alpha$ angles};\\
N_\beta^{ZM}&=2\times 2n=4n \quad\;\;\;\, \textrm{type $\beta$ angles}.
\end{aligned}
\end{equation}

\begin{figure}[h]
\centering
\includegraphics[scale=1]{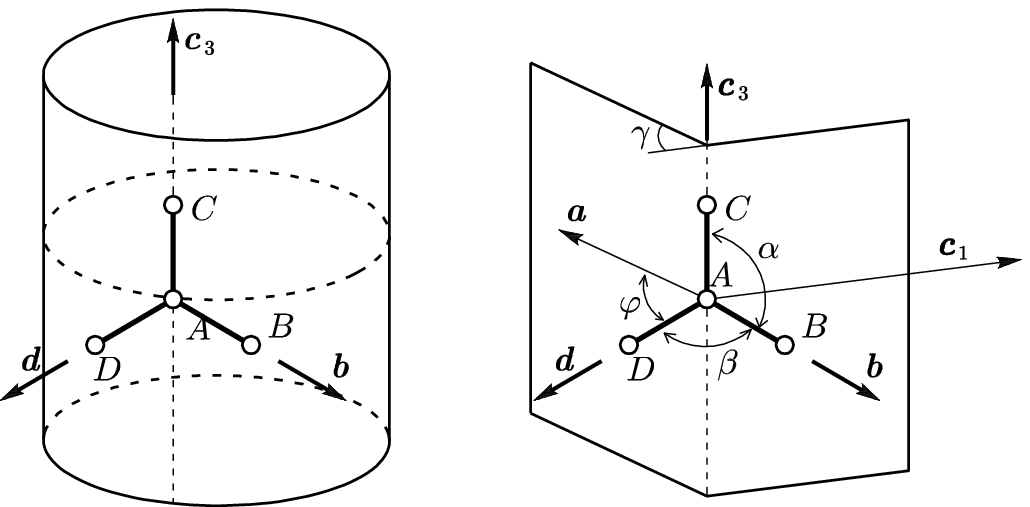}
\caption{The geometry of a ZBU: front view.} \label{framezig}
\end{figure}

\begin{figure}[h]
\centering
\includegraphics[scale=1]{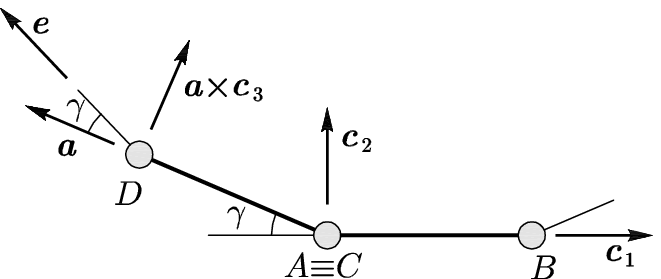}
\caption{The geometry of a ZBU: top-down view.} \label{zigzag_top}
\end{figure}

Figures \ref{framezig} and \ref{zigzag_top} illustrate a zigzag
bond unit (ZBU) and its geometry. 
We see that the three next neighbors of
the typical atom $A$, labeled $B$, $C$ and $D$, sit at the
following positions:
\begin{equation}\label{zigpos}
\begin{aligned}
&s^{-1}\pb_B:=\bb=\cos\varphi\,\cb_1-\sin\varphi\,\cb_3\\
&s^{-1}\pb_C:=\cb_3,\\
&s^{-1}\pb_D:=\db=\cos\varphi\,\ab-\sin\varphi\,\cb_3,
\end{aligned}
\end{equation}
with the unit vector  $\ab$ as in \eqref{avect}; note that $\alpha=\varphi+\pi/2$. It is not difficult to show that angles $\alpha,\beta$ and $\gamma$ are related as follows:
\begin{equation}\label{staqua}
\cos\beta=\cos^2\alpha-\sin^2\alpha\cos\gamma,
\end{equation}
whence
\begin{equation}\label{betaZ}
\beta=\widehat\beta^Z(n)=\arccos\left(\frac{1}{4}-\frac{3}{4}\cos\frac{\pi}{n}
\right).
\end{equation}
On repeating the considerations proposed  in Remark 1, and under the same circumstances, differentiation of \eqref{staqua} yields:
\begin{equation}\label{Zdiff}
\Delta\beta=\widehat\delta^Z(\alpha,\beta,\gamma)\Delta\alpha,\quad \widehat\delta^Z(\alpha,\beta,\gamma):= \frac{\sin
2\alpha}{\sin\beta}\,(1+\cos\gamma)=-\frac{\sqrt 3}{2\sin\beta}\,(1+\cos\gamma)\,.
\end{equation}
\remark \label{aZ}Here and henceforth, to lighten our notation, we write $\beta$ in the place of $\widehat\beta^Z(n)$ and $\widehat\beta^A(n)$; no confusion should arise, because the type of nanotube we deal with is always clear from the context.
\section{Nanoscopic DSM. Axisymmetric Equilibrium Problems}\label{bvps}
\begin{figure}[h]
\centering
\includegraphics[scale=0.75]{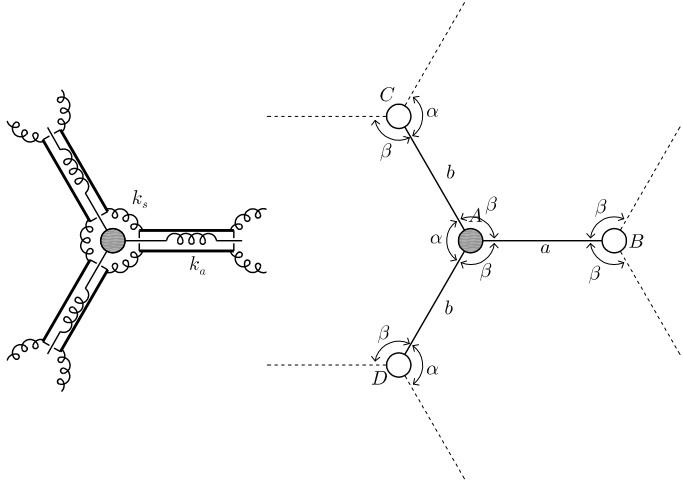}
\caption{A stick-and-spring caricature of an ABU.} \label{BU}
\end{figure}

\begin{figure}[h]
\centering
\includegraphics[scale=0.75]{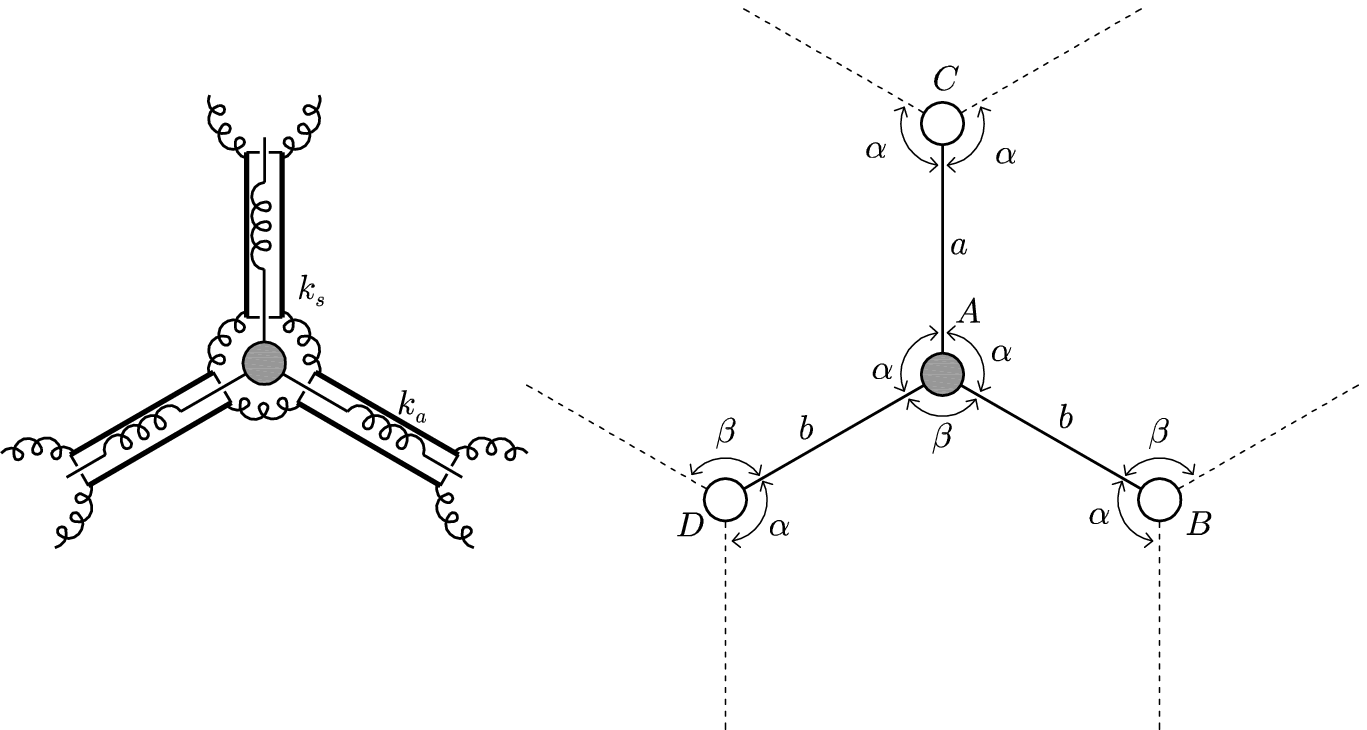}
\caption{A stick-and-spring caricature of a ZBU.} \label{mollezig}
\end{figure}

We view a meta-nanoscopic mechanical model of a SWCNT as consisting of pin-jointed rigid sticks and linearly elastic springs, of two types: (1) \emph{axial}, sensitive to changes in distance of the two C atoms sitting at the ends of the coaxial stick, of stiffness $k_a$; (2) \emph{spiral}, sensitive to changes in angle of the two sticks they are attached to, of stiffness $k_a$. Our mechanical model of an ABU looks as in Figure \ref{BU} (left);
the zigzag case is shown in Figure \ref{mollezig} (left).

In this section we apply the principles and methods of DSM to
solve explicitly a number of axisymmetric equilibrium problems for
SWCNTs, namely, \emph{axial traction and radial loading for both  A- and Z- CNTs}, and
\emph{torsion for ACNTs}. Problem by problem, different symmetries are presumed that simplify  considerably the solution process: to guess those symmetries correctly is at times not trivial; once they are detected, it is a matter of elementary vector algebra. For this reason, in each of the subsections here below, we relegate the ancillary computations to the Appendix.
\subsection{Axial traction of an ACNT}\label{ata}
\begin{figure}[h]
\centering
\includegraphics[scale=.8]{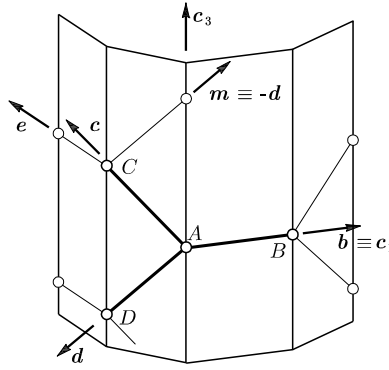}
\caption{An ABU and its adjacent sticks.} \label{EU_forz}
\end{figure}

This problem has been dealt with in \cite{CG,Li,SL}, for zigzag CNTs as well. Figure \ref{EU_forz}  is meant to give a helping hand to visualize an ABU and its immediately adjacent sticks when, on applying Euler's \emph{Cut Principle}, that ABU is required to be an equilibrated part of a nanotube subject to axial traction
and, accordingly, is regarded as acted upon by a balanced  system of forces and couples, due to its interactions with the adjacent sticks.
We assume that forces $\fb_B$, $\fb_C$, and $\fb_D$, such that
\begin{equation}\label{forbal}
\fb_B+\fb_C+\fb_D=\mathbf{0},
\end{equation}
are applied at points $B$, $C$, and $D$. In view of the peculiar symmetries of the axial traction problem,
we presume that
\begin{equation}\label{forze}
\begin{aligned}
&\fb_B\cdot\cb_3=0,\\&\fb_C\cdot\cb_3=-\fb_D\cdot\cb_3,\\&\fb_C\cdot\cb_\alpha=\fb_D\cdot\cb_\alpha=-\frac{1}{2}\,\fb_B\cdot\cb_\alpha\;\,(\alpha=1,2)
\end{aligned}
\end{equation}
(note that, with these presumptions, \eqref{forbal} is satisfied).

\begin{figure}[h]
\centering
\includegraphics[scale=.8]{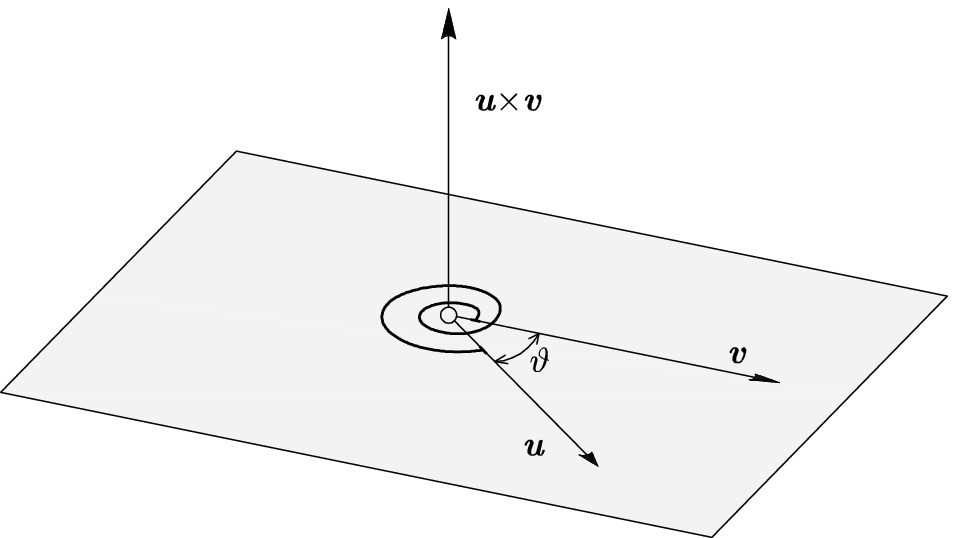}
\caption{The couple associated with a spiral spring.} \label{spi}
\end{figure}

In addition to forces, at both ends of each stick a number of couples are applied, which we distinguish into external and internal with reference to the Euler's cut which singled out the ABU. With a view toward writing moment balances, we stop and make our vectorial representation for couples unambiguous, with the help of Figure \ref{spi}.
The figure alludes to a linearly elastic spiral spring of stiffness $k_s$ and rest angle $\vartheta$, connecting two pin-jointed rigid sticks directed along  the unit vectors $\ub,\vb$. When the angle changes by $|\Delta\vartheta|$, the spring exerts a couple
$$
\taub=\tau\,\ub\times\vb,\;\,\tau=k_s\Delta\vartheta\,|\ub\times\vb|^{-1},
$$
on the stick along $\ub$ and a couple $\taub=-\tau\,\ub\times\vb$ on the other stick; thus, for $\Delta\vartheta<0$, a clockwise (counterclockwise) couple is exerted on the stick along $\ub$ ($\vb$). For example, upon loading the nanotube, the stick $AC$  has external couples $\bar\taub_\alpha$, $\bar\taub_\beta$ applied in $C$ and internal couples $\taub_\alpha$, $\taub_\beta$ applied in $A$, with
\begin{equation}\label{coup1}
\begin{aligned}
\bar\taub_\alpha&=\bar\tau_\alpha\,(-\cb)\times\mb,\;\,\bar\tau_\alpha=k_s\overline{\Delta\alpha}\, |\cb\times\mb|^{-1},\\
\bar\taub_\beta&=-\bar\tau_\beta\,\eb\times(-\cb),\;\,\bar\tau_\beta=k_s\overline{\Delta\beta} \,|\eb\times\cb|^{-1}
\end{aligned}
\end{equation}
and
\begin{equation}\label{coup2}
\begin{aligned}
\taub_\alpha&=\tau_\alpha\,\cb\times\db,\;\,\tau_\alpha=k_s\Delta\alpha \,|\cb\times\db|^{-1},\\
\taub_\beta&=-\tau_\beta\,\bb\times\cb,\;\,\tau_\beta=k_s\Delta\beta \,|\bb\times\cb|^{-1}.
\end{aligned}
\end{equation}

With a further application of the Cut Principle, the moments of the force and couple system acting on stick $AC$ are balanced with respect to the pole $A$ if it so happens that
\begin{equation}\label{eul}
\pb_C\times\fb_C+\taub_\alpha+\taub_\beta+\bar\taub_\alpha+\bar\taub_\beta=\mathbf{0}
\end{equation}
where we presume that
\begin{equation}\label{angvar}
\overline{\Delta\alpha}=\Delta\alpha,\quad \overline{\Delta\beta}=\Delta\beta
\end{equation}
(the problem's symmetries exonerate us from laying down the relation for stick $AD$, which would not add any relevant information). Likewise, moments of force and couples system acting on stick $AB$ are balanced with respect to $A$ provided that
$$
\pb_B\times\fb_B=\mathbf{0},
$$
whence
\begin{equation}\label{forba2}
\fb_B\cdot\cb_2=0.
\end{equation}
Algebraic manipulations to be found in Section \ref{app1} lead us to conclude that the changes in length of sticks of type $a$ and $b$ are:
\begin{equation}\label{allung}
\Delta a=\frac{\fb_B\cdot\bb}{k_a}=0, \quad \Delta b=\frac{\fb_C\cdot\cb}{k_a}=\sin\frac{\alpha}{2}\,f/k_a=\frac{\sqrt 3}{2}\,f/k_a,
\end{equation}
moreover, as to angle changes, we have:
\begin{equation}\label{deltavarib}
\begin{aligned}
&\Delta\alpha=\frac{1}{2}\,\cos\frac{\alpha}{2}\left(1+\frac{1}{2}\left(\frac{\tan\frac{\alpha}{2}}{\tan\beta}\right)^2  \right)^{-1}fs/k_s=\frac{1}{4}\left(1+\frac{3}{2\tan^2\beta}  \right)^{-1}fs/k_s\,,\\
&\Delta\beta=\frac{\sin\frac{\alpha}{2}}{4\tan\beta}\left(1+\frac{1}{2}\left(\frac{\tan\frac{\alpha}{2}}{\tan\beta}\right)^2  \right)^{-1}fs/k_s=\frac{\sqrt 3}{8\tan\beta}\left(1+\frac{3}{2\tan^2\beta}  \right)^{-1}fs/k_s\,,\\
\end{aligned}
\end{equation}
where angle $\beta$ depends on the chirality index $n$ as specified by \eqref{betaA} (recall Remark \ref{aZ}).

\subsection{Axial traction of a ZCNT}\label{atz}
\begin{figure}[h]
\centering
\includegraphics[scale=.8]{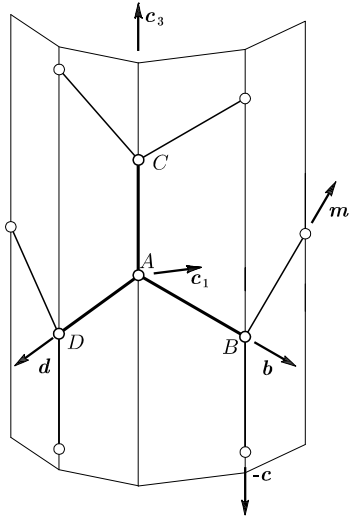}
\caption{A ZBU and its adjacent sticks.}
\label{EU_forze_Z}
\end{figure}

With reference to Figures \ref{EU_forze_Z} and \ref{zigzag_top},
let a system of forces satisfying \eqref{forbal} be applied at points $B,C,$ and $D$ of a typical ZBU, singled out in Euler's manner to assess its equilibrium conditions as a part of a zigzag nanotube subject to axial traction. The built-in symmetries require that:
\begin{equation}
\begin{aligned}\label{symzig}
&\fb_C=f\cb_3 ,
\\
 &\fb_B\cdot\cb_3=\fb_D\cdot\cb_3=-\frac{1}{2}\,\fb_C\cdot\cb_3=-\frac{1}{2}\,f,\\
&\fb_D\cdot\ab=\fb_B\cdot\cb_1,\quad \fb_D\cdot\ab\times\cb_3=\fb_B\cdot\cb_2.
\end{aligned}
\end{equation}
As to moment equilibrium, that of stick $AC$ with respect to $A$ is automatic, given the symmetries. Not so for stick $AB$: it must be that
\begin{equation}\label{Zeq}
\pb_B\times\fb_B+\taub_\alpha+\taub_\beta+\bar{\taub}_\alpha+\bar{\taub}_\beta=\mathbf{0},
\end{equation}
where
$$
\begin{aligned}
&\taub_\alpha=\tau_\alpha\bb\times\cb_3, \quad \tau_\alpha=k_s\Delta\alpha|\bb\times\cb_3|\\
&\bar{\taub}_\alpha=\bar{\tau}_\alpha(-\bb)\times(-\cb_3), \quad \bar\tau_\alpha=k_s\overline{\Delta\alpha}|\bb\times\cb_3|\\
&\taub_\beta=-\tau_\beta\,\db\times\bb, \quad \tau_\beta=k_s\Delta\beta|\db\times\bb|^{-1}\\
&\bar{\taub}_\beta=-\bar\tau_\beta\,\mb\times(-\bb), \quad
\tau_\beta=k_s\overline{\Delta\beta}|\bb\times\mb|^{-1}
\end{aligned}
$$
with $\bb,\db$ as in \eqref{zigpos} and
$$\mb=\sin\alpha(\cos\gamma\cb_1+\sin\gamma\cb_2)-\cos\alpha\cb_3.$$
Equation \eqref{Zeq} reads:
$$
\bb\times\Big(s\fb_B+2\tau_\alpha\cb_3+\tau_\beta(\db-\mb)  \Big)=0,
$$
where we presumed $\overline{\Delta\alpha}=\Delta\alpha$ and
$\overline{\Delta\beta}=\Delta\beta$.
After the algebraic manipulations confined in Section \ref{Aatz}, we conclude that
\begin{equation}\label{allung_zif}
\Delta a=\frac{\fb_C\cdot\cb}{k_a}=f/k_a, \quad \Delta
b=\frac{\fb_B\cdot\bb}{k_a}=-\frac{1}{2}\cos\alpha\,f/k_a=\frac{1}{4}\,f/k_a,
\end{equation}
and
\begin{equation}\label{deltavarib_zif}
\begin{aligned}
&\Delta\alpha=\frac{\sin\alpha}{4\left(1+2\frac{\tan^2\beta/2}{\tan^2\alpha}\right)}\,fs/k_s=\frac{3\sqrt{3}(1+\cos\beta)}{8(5+\cos\beta)}\,fs/k_s\\
&\Delta\beta=\frac{\cos\alpha\sin^2\alpha\sin\beta}{3-\cos\beta+\cos2\alpha\,(1-3\cos\beta)}\,fs/k_s=-\frac{3\sin\beta}{4(5+\cos\beta)}\,fs/k_s\,.
\end{aligned}
\end{equation}
\subsection{Torsion of an ACNT}\label{torz}
\begin{figure}[h]
\centering
\includegraphics[scale=.8]{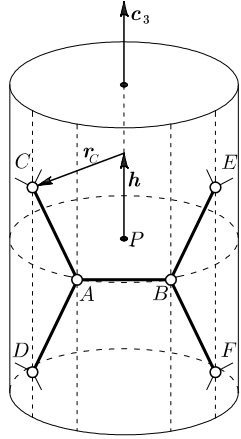}
\caption{A double ABU and its adjacent sticks: front view.}
\label{tors}
\end{figure}

This problem has been dealt with in \cite{Li,SL}. The Euler cut in  Figure \ref{tors}  individuates a double ABU, a five-stick complex that we require to be in equilibrium under a system of forces and couples applied at points $C,D,E$, and $F$.

\begin{figure}[h]
\centering
\includegraphics[scale=.8]{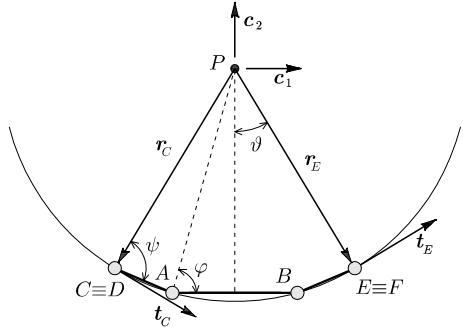}
\caption{A double ABU: top-down view.}
\label{tors1}
\end{figure}

As to forces, on guessing the problem's symmetries, we assume that
\begin{equation}\label{fortor}
\begin{aligned}
\fb_C&=h\tb_C+v\cb_3,\quad \fb_D=-h\tb_C+v\cb_3, \\ \fb_E&=h\tb_E-v\cb_3,\quad \fb_F=-h\tb_E-v\cb_3,
\end{aligned}
\end{equation}
with $\tb_C,\tb_E$ the unit vectors orthogonal to $\cb_3$ and tangent at points $C$ and $E$, respectively, to the cylinder on which all C atoms are sitting (see Figure \ref{tors1});
note that \eqref{fortor} implies that the resultant of this force system is null.

As to couples, we guess that angle changes are equal at points $C$ and $F$ and at points $D$ and $E$:
\begin{equation}\label{coptor}
\begin{aligned}
\Delta\alpha(C)&=\Delta\alpha(F),\quad \Delta\beta(C)=\Delta\beta(F);\\
\Delta\alpha(D)&=\Delta\alpha(E),\quad \Delta\beta(D)=\Delta\beta(E).
\end{aligned}
\end{equation}
and that, moreover, angle changes are opposite at points $C$ and $D$ and at points $E$ and $F$:
\begin{equation}\label{coptorbis}
\begin{aligned}
\Delta\alpha(C)&=-\Delta\alpha(D),\quad \Delta\beta(C)=-\Delta\beta(D);\\
\Delta\alpha(E)&=-\Delta\alpha(F),\quad \Delta\beta(E)=-\Delta\beta(F).
\end{aligned}
\end{equation}
Combining \eqref{coptor} and \eqref{coptorbis}, we arrive at the following set of symmetry-based assumptions:
\begin{equation}\label{coptorter}
\begin{aligned}
\Delta\alpha(C)&=\Delta\alpha(F)=-\Delta\alpha(E)=-\Delta\alpha(D),\\
\Delta\beta(C)&=\Delta\beta(F)=-\Delta\beta(E)=-\Delta\beta(D).
\end{aligned}
\end{equation}
The equilibrium of our stick complex is guaranteed if the resultant moment with respect to any chosen pole of the forces and couples applied at  points $C,D,E,F$ is null and if, moreover,  both the force and couple system acting on stick $AC$ and the force and couple system acting on stick $AD$ have null resultant moment with respect to point $A$.

The moment balance for the double ABU yields the scalar condition:
\begin{equation}\label{momentun}
(\rho_0\sin\vartheta)v-(k_s\cos\gamma)\Delta\alpha(C)+\Big(k_s\sin\frac{\alpha}{2}\frac{\cos 2\gamma}{\sin\beta}\Big)\Delta\beta(C)=-(s\sin\frac{\alpha}{2}\cos\vartheta)h\,.
\end{equation}
The moment balances for the $AC$ and $AD$ sticks furnish a symmetry condition that we had not been able to guess, namely,
\begin{equation}\label{delalf}
\Delta\alpha(A)=0,
\end{equation}
and  three scalar conditions:
and
\begin{equation}\label{somma}
\begin{aligned}
&k_s\frac{\sin\gamma}{\sin\beta}\big(\Delta\beta(C)+\Delta\beta^u(A)\big)=-\big(s\sin(\gamma-\vartheta)\big)h \,,\\
&\Big(s\cos\frac{\alpha}{2}\sin\gamma\Big)v-(k_s\sin\gamma)\Delta\alpha(C)+k_s\sin\frac{\alpha}{2}\frac{\sin2\gamma}{\sin\beta}\Delta\beta(C)=-\Big(s\sin\frac{\alpha}{2}\sin\vartheta\Big)h\,,\\
&\Big(s\cos\frac{\alpha}{2}\cos\gamma\Big)v-(k_s\cos\gamma)\Delta\alpha(C)+k_s\sin\frac{\alpha}{2}\frac{\cos2\gamma}{\sin\beta}\Delta\beta(C)+k_s\frac{\sin\frac{\alpha}{2}}{\sin\beta}\Delta\beta^u(A)\\&=-\Big(s\sin\frac{\alpha}{2}\cos\vartheta\Big)h\,,
\end{aligned}
\end{equation}
where $\Delta\beta^u(A)$ denotes the change in angle between the $AC$ and $AB$ sticks.
The developments necessary to arrive at these equations can be found in Section \ref{Atorz}.  The solution of the system of
\eqref{momentun} and \eqref{somma} is:
\begin{equation}\label{soluz}
\begin{aligned}
&v=-\sin\frac{\alpha}{2}\,\frac{\sin(\vartheta-\gamma)}{\sin\gamma}\,\xi\,h,\quad\textrm{with}\;\,\xi^{-1}:=\cos\frac{\alpha}{2}\cos\gamma-\frac{\rho_0}{s}\sin\vartheta,\\
&\Delta\alpha(C)=\sin\frac{\alpha}{2}\left(\cos\frac{\alpha}{2}\cos\vartheta - \frac{\rho_0}{s}\frac{\sin^2\vartheta}{\sin\gamma}\right)\,\xi\,hs/k_s,\\
&\Delta\beta(C)=0,\\
&\Delta\beta^u(A)=\sin\beta\,\frac{\sin(\vartheta-\gamma)}{\sin\gamma}\,hs/k_s.
\end{aligned}
\end{equation}
Note, in particular, $\eqref{soluz}_3$, another difficult-to-guess symmetry condition. Note also that it follows from the definition in \eqref{soluz}, with the use of \eqref{rozzo}, \eqref{cos}, and \eqref{teta}, that $\xi=-2$.
Thus, the first two of \eqref{soluz} can be written in the following simpler form:
\begin{equation}\label{soluz2}
\begin{aligned}
&v=2\sin\frac{\alpha}{2}\,\frac{\sin(\vartheta-\gamma)}{\sin\gamma}\,h,\\
&\Delta\alpha(C)=2\sin\alpha\,\frac{\sin(\vartheta-\gamma)+\sin\vartheta}{\sin\gamma}\,hs/k_s.
\end{aligned}
\end{equation}

It follows from the first relations in \eqref{fortor}, \eqref{titti}, and \eqref{soluz}, that the axial force in the sticks $AC$ and $AD$ is:

$$
\fb_C\cdot\cb=: N(AC)=-N(AD)=-\left(
\cos\frac{\alpha}{2}\cos(\vartheta-\gamma)-2\sin^2\frac{\alpha}{2}\,\frac{\sin(\vartheta-\gamma)}{\sin\gamma}
\right)\,h\,;
$$
thus, their axial deformation is:
\begin{equation}\label{alltor}
\Delta b(AC)=-\Delta b(AD)=-\left(
\cos\frac{\alpha}{2}\cos(\vartheta-\gamma)-2\sin^2\frac{\alpha}{2}\,\frac{\sin(\vartheta-\gamma)}{\sin\gamma}
\right)\,h/k_a.
\end{equation}
Moreover, the axial force in stick $AB$ turns out to be null, so therefore $$\Delta a=0.$$

\subsection{Radial loading of an ACNT}\label{rada}
\begin{figure}[h]
\centering
\includegraphics[scale=.8]{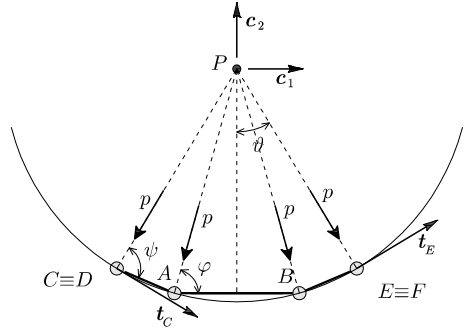}
\caption{A double ABU subject to uniform radial forces: top-down
view.} \label{pres}
\end{figure}

Figure \ref{pres} features a top view of a double ABU
subject to radial forces of common intensity $p$,
a load condition that we later assimilate to the radial pressure problem for a cylindrical shell.
 The front view is the same as in Figure \ref{tors}. We presume the following symmetries, for the applied forces consequential to our Euler cut:
\begin{equation}\label{forzepres}
\begin{aligned}
&\fb_C=\fb_D,\quad \fb_E=\fb_F,\\&\fb_C\cdot\cb_1=-\fb_E\cdot\cb_1,\\&\fb_C\cdot\cb_2=\fb_E\cdot\cb_2,\\&\fb_C\cdot\cb_3=\fb_E\cdot\cb_3=0,
\end{aligned}
\end{equation}
and for the angle changes:
\begin{equation}\label{coppre}
\Delta\alpha(A)=\Delta\alpha(C)=:\Delta\alpha,\quad \Delta\beta(A)=\Delta\beta(C)=:\Delta\beta,
\end{equation}
where $\Delta\alpha,\Delta\beta$ satisfy \eqref{deltavari}.

Given \eqref{forzepres}, and given that
\[
\fb_A+\fb_B=-2\sin\varphi\, p\,\cb_2,
\]
the resultant of the force system is null if
\[
-p\sin\varphi\,\cb_2+\fb_C+\fb_E=\mathbf{0},
\]
i.e., if, in addition to the second of \eqref{forzepres},
\begin{equation}\label{effeC}
\fb_C\cdot\cb_2=\frac{\sin\varphi}{2}\,p.
\end{equation}
Only one more balance equation is needed, and we choose it to express the rotation equilibrium of stick $AC$ with respect to
point $A$:
\begin{equation}\label{eqpres}
\pb_C\times\fb_C+\taub_\alpha+\taub_\beta+\bar\taub_\alpha+\bar\taub_\beta=\mathbf{0},
\end{equation}
where, with the use of \eqref{coppre},
\begin{equation}\label{tauvarie}
\begin{aligned}
&\taub_\alpha=\tau_\alpha\,\cb\times\db, \quad
\tau_\alpha=k_s\Delta\alpha\,|\cb\times\db|^{-1},\\
&\taub_\beta=-\tau_\beta\, \bb\times\cb, \quad
\tau_\beta=k_s\Delta\beta \,|\cb\times\bb|^{-1},\\
&\bar\taub_\alpha=\bar\tau_\alpha\,(-\cb)\times(-\db), \quad
\bar\tau_\alpha=k_s\Delta\alpha\,|\cb\times\db|^{-1},\\
&\bar\taub_\beta=-\bar\tau_\beta\,\eb\times(-\cb), \quad
\bar\tau_\beta=k_s\Delta\beta\,|\cb\times\eb|^{-1}.
\end{aligned}
\end{equation}

The algebraic manipulations in Section \eqref{Arada} allow us to conclude that
\begin{equation}\label{prealfa}
\Delta\alpha=-\frac{\sin\alpha\sin\varphi}{2\left(4\cos\frac{\alpha}{2}\sin\gamma+\widehat\delta^A(\alpha,\beta,\gamma)\sin\alpha\sin^{-1}\beta\sin 2\gamma\right)}\,ps/k_s,
\end{equation}
where, we recall,
\[
\quad \widehat\delta^A(\alpha,\beta,\gamma)=\frac{\sin\frac{\alpha}{2}}{2\sin\beta}\,\cos\gamma;
\]
as to the axial deformation, we obtain:
\begin{equation}\label{prea}
\Delta a=\big(\widehat\delta^A(\alpha,\beta,\gamma)\sin\varphi\big)p/k_a,
\end{equation}
and
\begin{equation}\label{preb}
\Delta
b=\frac{1}{2}\Big(\cos\frac{\alpha}{2}\left(\sin\gamma+\cot\gamma\cos\gamma\right)\sin\varphi\Big)p/k_a\,.
\end{equation}

\subsection{Radial loading of a ZCNT}\label{325}
\begin{figure}[h]
\centering
\includegraphics[scale=.8]{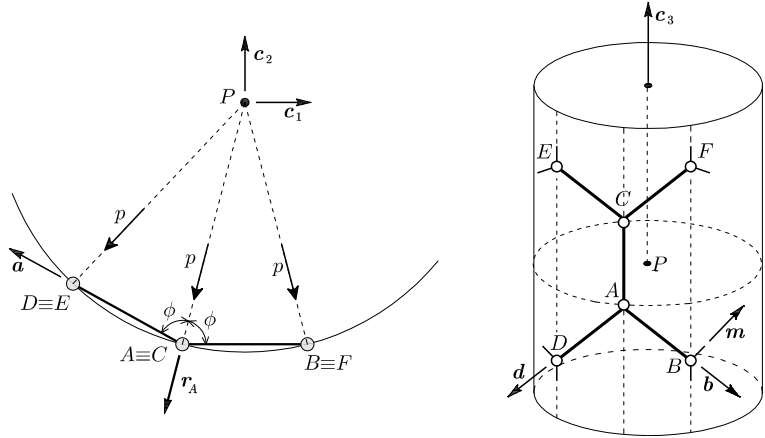}
\caption{A double ZBU subject to uniform radial forces: top-down and
front views.} \label{preszig}
\end{figure}

%
Consider a double ZBU, subject to applied radial forces of common intensity $p$ (Figure \ref{preszig}), and imagine it  isolated from the CNT it belongs to by means of an Euler cut.

We see that:
\begin{equation}\label{fo1}
\fb_C=\fb_A=p\,\rb_A,\quad \textrm{with}\quad \rb_A:=-\left(\sin\frac{\gamma}{2}\,\cb_1+\cos\frac{\gamma}{2}\,\cb_2\right)
\end{equation}%
(with reference to Figure \ref{preszig}-left, note that $\phi=(\pi-\gamma)/2$).
The force symmetries we presume are:
\[
\begin{aligned}\label{fo2}
\fb_D=\fb_E,\quad \fb_B=\fb_F, \quad\fb_D\cdot\rb_A=\fb_B\cdot\rb_A,\quad \fb_D\cdot\rb_A^\perp=-\fb_B\cdot\rb_A^\perp,
\end{aligned}
\]
where $\,\rb_A^\perp:=\rb_A\times\cb_3$;
in addition, we presume that
\begin{equation}\label{fo3}
\fb_D\cdot\cb_3=\fb_B\cdot\cb_3=0\,.
\end{equation}
Under these presumptions,  only two balance equations are needed for equilibrium: the resultant of all forces applied at points $A,B,C,D,E,$ and $D$ and the resultant moment with respect to $A$ of all forces and couples acting on stick $AB$ must both be null (see Subsection \ref{325} of the Appendix).

In view of \eqref{fo1}, \eqref{fo2}, and \eqref{fo3}, the force balance equation reads:
\[
\fb_B+\fb_D+p\,\rb_A=\mathbf{0},
\]
and is equivalent to its only scalar consequence:
\begin{equation}
(\fb_B+\fb_D)\cdot\rb_A+p=0,
\end{equation}
or rather,
\begin{equation}\label{condz}
\sin{\frac{\gamma}{2}}\fb_B\cdot\cb_1+\cos\frac{\gamma}{2}\fb_B\cdot\cb_2=\frac{1}{2}\,p
\,.
\end{equation}

The moment balance for stick $AB$ reads:
\begin{equation}\label{eqpresz}
\pb_B\times\fb_B+\taub_\alpha+\taub_\beta+\bar\taub_\alpha+\bar\taub_\beta=\mathbf{0},
\end{equation}
where,
\begin{equation}\label{tauvariez}
\begin{aligned}
&\taub_\alpha=\tau_\alpha\,\bb\times\cb_3, \quad
\tau_\alpha=k_s\Delta\alpha|\bb\times\cb_3|^{-1},\\
&\taub_\beta=-\tau_\beta\,\db\times\bb, \quad
\tau_\beta=k_s\Delta\beta|\db\times\bb|^{-1},\\
&\bar\taub_\alpha=\bar\tau_\alpha\,\bb\times\cb_3, \quad
\bar\tau_\alpha=k_s\Delta\alpha|\bb\times\cb_3|^{-1},\\
&\bar\taub_\beta=-\bar\tau_\beta\,\mb\times(-\bb), \quad
\bar\tau_\beta=k_s\Delta\beta|\db\times\bb|^{-1}.\\
\end{aligned}
\end{equation}

The developments leading to establishing that
\begin{equation}\label{prealfaz}
\left\{\begin{array}{c}\Delta\alpha=\displaystyle{-\frac{1}{4\sin\frac{\gamma}{2}\left(2-\frac{\widehat\delta^Z(\alpha,\beta,\gamma)}{\sin\beta}(\sin\alpha\cos\gamma+\sin\alpha)\right)}\frac{ps}{k_s},}\qquad\qquad\;\\
 \Delta\beta=\widehat\delta^Z(\alpha,\beta,\gamma)\Delta\alpha,\quad \displaystyle{\widehat\delta^Z(\alpha,\beta,\gamma)=-\frac{\sqrt 3}{2\sin\beta}\,(1+\cos\gamma)\,,\qquad\;\,}
 \end{array}\right.
\end{equation}
$$
\Delta a=0,
$$
and
\begin{equation}\label{prebz}
\Delta b=\frac{\sin\alpha}{2\sin\frac{\gamma}{2}}\frac{p}{k_a},
\end{equation}
are found in Section \ref{A325}.

\subsection{Energetics}
The elastic energy stored in a deformation can be evaluated in a number of alternative ways.
 We find it convenient to do it with reference to bond units.
\subsubsection{The energy stored in A- and Z- CNTs under axial traction}\label{subsuben}
 With reference to Figure \ref{BU},
 the total energy stored in an ABU turns out to be:
\begin{equation}\label{armen}
\begin{aligned}
U^{ABU}_{Ax}&=\frac{1}{2}k_a\Big((\Delta a)^2+2(\Delta b)^2
\Big)+\frac{1}{2}k_s\Big((\Delta\alpha)^2+2(\Delta\beta)^2+2\times\frac{1}{2}(\Delta\alpha)^2+4\times\frac{1}{2}(\Delta\beta)^2\Big)
\\
&=k_a(\Delta b)^2
+k_s\Big((\Delta\alpha)^2+2(\Delta\beta)^2\Big),
\end{aligned}
\end{equation}
where $\Delta b$ is given by $\eqref{allung}_2$ and $\Delta\alpha,\Delta\beta$ by \eqref{deltavarib}; for a ZBU (Figure \ref{mollezig}), we find:
\begin{equation}\label{zigen}
\begin{aligned} U^{ZBU}_{Ax}&=\frac{1}{2}k_a\Big((\Delta a)^2+2(\Delta
b)^2
\Big)+\frac{1}{2}k_s\Big(2(\Delta\alpha)^2+(\Delta\beta)^2+4\times\frac{1}{2}(\Delta\alpha)^2+2\times\frac{1}{2}(\Delta\beta)^2\Big)
\\
&=\frac{1}{2}k_a\Big((\Delta a)^2+2(\Delta b)^2
\Big)+k_s\Big(2(\Delta\alpha)^2+(\Delta\beta)^2\Big),
\end{aligned}
\end{equation}
with $\Delta a,\Delta b$ given by \eqref{allung_zif} and $\Delta\alpha,\Delta \beta$ by \eqref{deltavarib_zif}.
 There are $4n$ ABUs and $2n$ ZBUs in a module, so that the energy
stored per module is:
\begin{equation}\label{moden}
U^{AM}_{Ax}=4n\, U^{ABU}_{Ax},\quad U^{ZM}_{Ax}=2n\, U^{ZBU}_{Ax}.
\end{equation}
\subsubsection{The energy stored in a twisted armchair CNT}
In this instance, the energy stored in an ABU is:
\begin{equation}\label{entor}
\begin{aligned}
U^{ABU}_{To}&=\frac{1}{2}k_a\Big[\big(\Delta a\big)^2+2\big(\Delta
b(C)\big)^2+2\big(\Delta b(D) \big)^2
\Big]+\frac{1}{2}k_s\Big[2\big(\Delta\alpha
\big)^2+2\big(\Delta\beta^u(A)\big)^2+\\
&+2\big(\Delta\beta^d(A)\big)^2+\big(\Delta \alpha(C)
\big)^2+\big(\Delta \alpha(D) \big)^2+\big(\Delta \beta(C)
\big)^2+\big(\Delta \beta(D) \big)^2\Big]=\\
&= 2k_a\big(\Delta
b(AC)\big)^2+k_s\Big[2\big(\Delta\beta^u(A)\big)^2+\big(\Delta
\alpha(C) \big)^2\Big],
\end{aligned}
\end{equation}
where $\Delta\beta^u(A)$ is given by $\eqref{soluz}_{4}$, $\Delta \alpha(C)$ by $\eqref{soluz2}_{2}$, and $\Delta b(AC)$  by \eqref{alltor}.
\subsubsection{The energy stored in A- and Z- CNTs subject to radial loading}
The energy stored in an ABU is:
\begin{equation}\label{enrad}
U^{ABU}_{Rd}=\frac{1}{2}k_a\Big(\big(\Delta a\big)^2+2\big(\Delta
b\big)^2
\Big)+k_s\big(1+2(\widehat\delta^A(\alpha,\beta,\gamma))^2\big)\big(\Delta\alpha
\big)^2,
\end{equation}
where $\Delta a,\Delta b$, and $\Delta \alpha$, are given, respectively, by \eqref{prea}, \eqref{preb}, and \eqref{prealfa}. For a ZBU,
the stored energy is:
\begin{equation}\label{enradz}
U^{ZBU}_{Rd}=k_a(\Delta
b)^2+(2+(\widehat\delta^Z(\alpha,\beta,\gamma))^2)k_s\big(\Delta\alpha\big)^2,
\end{equation}
where $\Delta b$ and $\Delta \alpha$ are given by, respectively, \eqref{prebz} and $\eqref{prealfaz}_1$.
\subsection{Slenderness and thinness}\label{slen}
We here discuss two important properties of a SWCNT, its slenderness and its thinness, within the context  of DSM.

For $N$ the number of C atoms at our disposal (say, the largest number of atoms we can treat in a numerical simulation), those atoms can be arranged in a nanotube consisting of $\textrm{int}(N/N_M)$ modules, where $\textrm{int}(\cdot)$ denotes the integer-part mapping. For $l_M$ the length of a module, we have the following formula for the length $L$ of the nanotube in question:
\[
L\approx \textrm{int}(N/N_M)\,l_M=\left\{\begin{array}{c} 2\sqrt{3}\,\textrm{int}(N/8n)\,s \;\;
(\textrm{armchair})\\\\ \!\!\!{3}\,\textrm{int}(N/4n)\,s
\;\; (\textrm{zigzag})\end{array}\right.\!\!\!.
\]
Consequently, that nanotube's \emph{aspect ratio} -- that is, its diameter/length ratio -- is:
\begin{equation}\label{slend}
\frac{2\rho_0}{L}\approx\left\{\begin{array}{c} \frac{\sqrt 3}{2\pi}\,\frac{n}{\textrm{int}(N/8n)} \quad
(\textrm{armchair})\\\\ \frac{1}{\sqrt{3}\,\pi}\,\frac{n}{\textrm{int}(N/4n)}
\;\; (\textrm{zigzag})\end{array}\right.\!\!\!.
\end{equation}
where  use has been made of \eqref{rozzo} and \eqref{roz}. Therefore, a nanotube is \emph{slender} if, roughly speaking, $n^2\ll N$.
Consistently, when matching
theoretical predictions with computational results, convenient
simulations should concern arrays of some $N=10^3n^2$ atoms or
more, as a glance to \eqref{slend} indicates.

The chirality index $n$ has to do not only with the slenderness of A- and Z- CNTs, but also with their \emph{thinness}. A purely geometrical notion of thinness is easy to state within the framework of CSM: given a cylindrical shell of
diameter $2(\rho_o+\varepsilon)$ and length $2l$, we term it \emph{thin}   if
$\varepsilon/\rho_o\ll 1$, that is to say, if its thickness-to-diameter ratio is small.\footnote{See \cite{PPGF} for a discussion of the non-geometric ingredients that a notion of thinness may contain.}  Thickness, however, is an ill-defined notion for CNTs, especially when
they are single-wall. This is why, in our opinion, a purely geometrical discrete notion of thinness makes no sense for CNTs. Be it as it may, on recalling relation \eqref{raggio} between radius and chirality, we can estimate as follows the `geometrical thinness' of a $(n,m)-$SWNT of arbitrary chirality and radius $\rho_0$:
\begin{equation}\label{thin}
\frac{s}{2\,\widehat\rho_0(n,m)}=\frac{\pi}{n\sqrt{3\big(1+m/n+(m/n)^2\big)}} =\left\{\begin{array}{c} \frac{\pi}{3}\,n^{-1} \quad (\textrm{armchair})\\\\\!\!\!\!\frac{\pi}{\sqrt 3}\,n^{-1} \;\; (\textrm{zigzag})\end{array}\right.\!\!\!.
\end{equation}
Thus, for an ACNT to be thin, the chirality index $n$ must be larger than, say, 10 or more, a rather rare occurence. We shall see in Section \ref{costpar} that a more appropriate, not purely geometrical, notion of thinness does yield a decreasing, but less pronounced, dependence on $n$: \emph{shells mimicking the mechanical behavior  of SWCNTs are generally not thin!}

\section{Nanoscopic CSM. A SWCNT-Oriented Shell Theory }\label{macro}
\begin{figure}[h]
\centering
\includegraphics[scale=0.7]{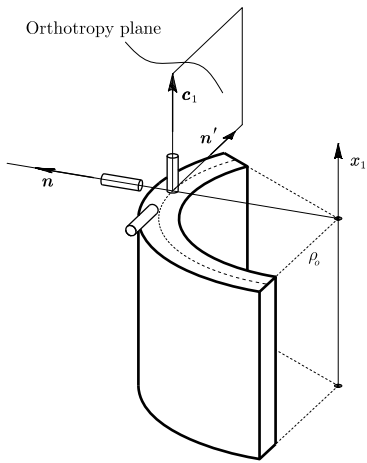}
\caption{The geometrical elements relevant to describe the
material response of CNTs modeled as shells.} \label{ortho}
\end{figure}

In the majority of papers where CNTs are regarded
macroscopically as shells,  textbook theories induced from classic
three-dimensional \emph{isotropic} elasticity are used;
consequently, the shell response is characterized in terms of
two elastic moduli.\footnote{There are, however, noticeable exceptions to this oversimplifying practice, such as \cite{Ru} and \cite{Robby}.}  On the basis of the DSM modeling at the nanoscale
of armchair and zigzag CNTs we here use, we think it better to view all CNTs, be they single- or
multi-wall and whatever their chirality, as \emph{orthotropic}
cylindrical shells whose midsurface  has a tangent plane
coinciding with the orthotropy plane, so that the shell geometry
agrees point-wise with the geometry intrinsic to the chosen type
of material response (see Fig. \ref{ortho},
where the three little
cylinders suggest what probes one should cut out of the shell body
in order to determine its material moduli). Such a shell theory has been fully developed in \cite{Fa2}; we here employ the simplest version fitting armchair and zigzag CNTs.
%
%
\subsection{Displacement and strains}
\begin{figure}[h]
\centering
\includegraphics[scale=1]{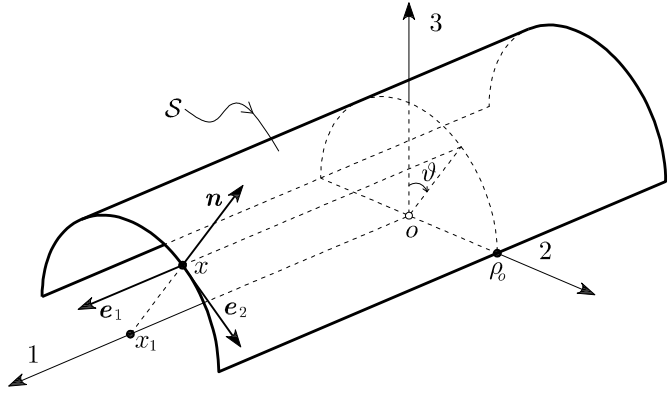}
\caption{A portion of the model surface $\mathcal S$ of a right cylindrical
shell.}
\label{cyl}
\end{figure}

With reference to Fig. \ref{cyl}, given that $(x_1,\vartheta,\zeta)\in(-l,+l)\times(0,2\pi)\times(-\varepsilon,+\varepsilon)$, the typical shell we consider as \emph{length} $2\,l$, \emph{wall thickness} $2\,\varepsilon$, and \emph{aspect ratio} $\rho_o/l$; we call it \emph{slender} if $\rho_o/l\ll 1$.
In the theory we import from \cite{Fa2}, displacement fields inducing thickness changes and thickness shearing are excluded a priori. In addition, for our present purposes, consideration of axisymmetric solutions to axisymmetric boundary value problems suffices. These solutions are all independent of the circumferential coordinate $\vartheta$, and have the following form in physical components $u{\<{i}}$:
\begin{equation}\label{cilcil}
u{\<1}=a{\<1}-\zeta w', \quad u{\<2}=\left(1+\frac{\zeta}{\rho_o}
\right)a{\<2}, \quad u{\<3}=w,
\end{equation}
where a prime denotes differentiation with respect to $x_1$, the
only space variable from which all of the parameter fields
$a{\<1},a{\<2}$, and $w$, may depend.
With the use of the
linear strain measure
\begin{equation}\label{strain}
\Eb(\ub)=\frac{1}{2}(\nabla\ub+\nabla\ub^T),
\end{equation}
it is not difficult to see that
$$\Eb(\ub)\nb=\mathbf{0},$$
i.e., that a
shell of the type under study cannot change its thickness and that its fibers
orthogonal to the referential midsurface $\Sc$ must remain
orthogonal to it after any admissible deformation. The not identically null strain components are:
\begin{equation}\label{deff}
\begin{aligned}
E{\<{11}}&=a{\<1}'-\zeta w''\,,\\E{\<{12}}&=\frac{1}{2}\left( 1+\frac{\zeta}{\rho_o}
\right)a{\<2}', \\E_{\<{22}}&=\left(\rho_o\left(
1+\frac{\zeta}{\rho_o} \right) \right)^{-1}\!\!w
\end{aligned}
\end{equation}
(needless to say, $E{\<{12}}=E{\<{21}}$).
On defining the \emph{cross-section strain measure}:
\begin{equation}\label{sects}
\overline{\Eb}(x_1):=\frac{1}{2\varepsilon}\int_{-\varepsilon}^{+\varepsilon}
\left(1+\frac{\zeta}{\rho_o} \right)\, \Eb(x_1,\zeta),
\end{equation}
we have from \eqref{deff} that
\begin{equation}\label{compE}
\overline{E}{\<{11}}=a{\<{1}}',\quad\overline{E}{\<{12}}=\overline{E}{\<{21}}=\frac{1}{2}\,a{\<{2}}',\quad\overline{E}{\<{22}}=\rho_o^{-1}w\,.
\end{equation}
\subsection{Constitutive parameters}
The mechanical response of the material comprising the cylindrical shells can be expressed in terms of
five elastic moduli: $E_1, E_2, \nu_{12},
\nu_{21}, G$; the first two are Young-like moduli, the second two being Poisson-like, and the fifth being a shear modulus associated with the $(\eb_1,\eb_2)$ pair of directions; the first four are not independent, because it must be that
\begin{equation}\label{ennu}
\frac{E_1}{E_2}=\frac{\nu_{12}}{\nu_{21}}\,.
\end{equation}
\subsection{Axisymmetric equilibria}
Here below we record the equilibrium solutions of the axial traction, torsion, and pressure, problems;  once again, the reader is referred to \cite{Fa2} for details.
\subsubsection{Axial Traction}
Suppose the only loading is a
distribution of end tractions equivalent to two mutually
balancing axial forces of magnitude
\begin{equation}\label{axtraz}
F=(2\pi\rho_o)p,\quad \textrm{with}\;\,p=O(\varepsilon).
\end{equation}
One finds that $a{\<{2}}\equiv 0$ and that, for a slender shell,
\begin{equation}\label{at1}
\begin{aligned}
w(x_1)&=-\rho_o\,\nu_{12}\,\delta(\varepsilon/\rho_o)\,\frac{\varepsilon
^{-1}p}{2E_1}, \\ a{\<{1}}(x_1)&=
\big(1-\nu_{12}\nu_{21}\left(1-
\delta(\varepsilon/\rho_o)\right)\big)\frac{\varepsilon^{-1}p}{2E_1}\,x_1,
\end{aligned}
\end{equation}
where
\begin{equation}\label{deltaeps}
 \delta(\varepsilon/\rho_o):=\frac{1-\nu_{12}\nu_{21}}{\frac{1}{2\frac{\varepsilon}{\rho_o}}\log\frac{1+\frac{\varepsilon}{\rho_o}}{1-\frac{\varepsilon}{\rho_o}}-\nu_{12}\nu_{21}};
 \end{equation}
note for later use that
\[
\lim_{\varepsilon/\rho_o\rightarrow 0}\,\frac{1}{2\frac{\varepsilon}{\rho_o}}\log\frac{1+\frac{\varepsilon}{\rho_o}}{1-\frac{\varepsilon}{\rho_o}}\,=1,\quad\textrm{whence}\quad\lim_{\varepsilon/\rho_o\rightarrow 0}\delta(\varepsilon/\rho_o)=1.
\]
\subsubsection{Torsion}
When the only applied load is a distribution of end tractions
statically equivalent to two mutually balancing torques of
magnitude
\begin{equation}\label{tigrande}
T=(2\pi\rho_o^2)t, \quad\textrm{with}\;\, t=O(\varepsilon),
\end{equation}
the axial and radial displacements  $a{\<{1}}$ and $w$ vanish
identically, whereas
\begin{equation}\label{to}
a{\<{2}}(x_1)=\frac{1}{1+\frac{\varepsilon^2}{\rho_o^2}}\frac{\varepsilon^{-1}t}{2G}\,x_1.
\end{equation}
As shown in \cite{Fa2}, these results hold for whatever the value of the aspect ratio.
\subsubsection{Radial pressure}
When the slender shell under study is subject to a uniform radial
pressure $\varpi=O(\varepsilon)$, all the other applied
loads being null, one again finds that $a{\<{2}}\equiv 0$. Moreover, for a slender shell,\footnote{See \cite{Fa2} for a full treatment of the traction and pressure problems for shells of arbitrary aspect ratio.}
\begin{equation}\label{w_presKL}
w(x_1)= \rho_o^2\,\delta(\varepsilon/\rho_o)
\frac{\varepsilon^{-1}\varpi}{2E_2}\quad\textrm{and}\quad a{\<{1}}=-\nu_{21}\,\delta(\varepsilon/\rho_o)\,\frac{\rho_o\eps^{-1}\varpi}{2E_2}\,x_1.
\end{equation}

\section{The Effective Parameters of CNT-like Shells}\label{effpar}
No matter whether we model a SWCNT as a shell or as a stick-and-spring complex, we may regard it as a \emph{cylindrical probe}. Two parameters characterize the mechanical response of a probe in an axial traction experiment, and one in torsion; their verbal definitions are, respectively,
\begin{equation}\label{para12}
s_{Ax}:=\frac{{\rm axial\,\,load}}{{\rm
axial\,\, deformation}},\quad
 \nu_{Ax}:=-\frac{{\rm radial\,\,deformation}}{{\rm
axial\,\,
deformation}},
\end{equation}
and
\begin{equation}\label{para3}
s_{T\!o}=\frac{{\rm axial \,\,torque}}{{\rm axial \,\,
twist}};
\end{equation}
$s_{Ax}$ and $s_{To}$ are stiffness measures, $\nu_{Ax}$ is a Poisson-like
modulus of transverse contraction. Two additional parameters are
needed to characterize the response of a SWCNT to radial loading,
namely, the specific stiffness and Poisson-like moduli
\[
s_{Rd}:=\frac{\rm radial\,\,load}{\rm radial\,\,deformation},\quad
 \nu_{Rd}:=-\frac{{\rm axial\,\,deformation}}{{\rm
radial\,\,
deformation}}.
\]

In this section we derive mathematical expressions for these five
parameters within the frameworks of both continuous and discrete
structure mechanics. The CSM expressions involve $E_1, E_2,
\nu_{12}, \nu_{21}$, and $G$, the five elastic moduli of the shell
theory we adopt, as well as two geometric parameters, $\varepsilon$ and $\rho_o$, the shell's thickness and
model-surface radius. The DSM
expressions depend on the nanoscopic spring-stiffness moduli $k_a$
and $k_s$, as well as on $s$, the C-C bond length, and $\rho_0$,
the radius of the cylinder on which the C atoms arrange themselves
(both for A- and Z- CNTs, $\rho_0\propto ns$, recall \eqref{rozzo} and \eqref{roz}). We aim
to evaluate the constitutive and geometric parameters of our nanoscopically informed theory of CNT-like shells solely  in
terms of $k_a,k_s,s$, and $n$. Our simple method consists in equating as many as needed of the continuous and discrete expressions we derive in the next two subsections for the various stiffnesses and contraction moduli listed above.

\subsection{`Continuous' parameters}\label{61}
When we regard a slender cylindrical shell as a probe subject to an axial load $F$, we set:
%
\begin{equation}\label{rigtrap}
s_{Ax}^{(c)}:=\frac{F}{\overline{E}{\<{11}}}\quad\textrm{and}\quad \nu^{(c)}_{Ax}:=-\frac{{\overline E}{\<{22}}}{{\overline
E}{\<{11}}}
\end{equation}
for the effective \emph{axial stiffness} and \emph{axial contraction modulus}. With the use of \eqref{compE} and \eqref{at1}, we then find:
\begin{equation}\label{rigtra}
s_{Ax}^{(c)}=\frac{1}{1-\nu_{12}\nu_{21}\left(1-
 \delta(\varepsilon/\rho_o)\right)}\,E_1A(\eps),\quad \nu^{(c)}_{Ax}=\frac{\delta(\varepsilon/\rho_o)}{1-\nu_{12}\nu_{21}\left(1-
 \delta(\varepsilon/\rho_o)\right)}\,\nu_{12},
\end{equation}
where $A(\varepsilon):=4\pi\rho_o\varepsilon$ is the area of the shell's
cross-section.

For a twisted shell of arbitrary slenderness, we set:
\begin{equation}\label{rigtors1}
s_{T\!o}^{(c)}:=\frac{T}{\Theta},\quad \Theta:=
\rho_o^{-1}a{\<{2}}^\prime,
\end{equation}
where $\Theta$ denotes the change in cross-section rotation angle per unit length caused by the application of the axial torque $T$.  Accordingly, with the use of \eqref{to}, we find for the effective \emph{torsional stiffness}:
\begin{equation}\label{rigtors}
s_{T\!o}^{(c)}=GJ(\varepsilon),\quad
J(\varepsilon):=4\pi\rho_o^3\varepsilon\Big(1+\frac{\varepsilon^2}{\rho_o^2}\Big)^{-1},
\end{equation}
where $J(\varepsilon)$ is the polar inertia moment of the cross section.\footnote{Recall that \eqref{to}, and hence \eqref{rigtors}, holds whatever the
slenderness of the shell under consideration.}

In case of a uniform radial pressure $\varpi$, we define
for our slender shell-like probe an effective \emph{radial strain}
$\rho_o^{-1}w$, in terms of which we set
\[s_{Rd}^{(c)}:=\frac{\varpi}{\rho_o^{-1}w}\quad \textrm{and}\quad \nu^{(c)}_{Rd}:=-\,\frac{E{\<{11}}}{\rho_o^{-1}w}
\]
for, respectively, the effective \emph{radial stiffness} and the effective \emph{radial contraction modulus}. With these definitions, relations \eqref{w_presKL} yield:
\begin{equation}\label{radsti}
s_{Rd}^{(c)}=\frac{2 E_2\,\eps/\rho_o}{\delta(\varepsilon/\rho_o)}\,,\quad
\nu^{(c)}_{Rd}=\nu_{21}.
\end{equation}
\vskip 6pt \remark The dimensions of the stiffness parameters are:
\[
\textrm{dim}(s_{Ax}^{(c)})=\textrm{force},\quad \textrm{dim}(s_{To}^{(c)})=\textrm{force}\times(\textrm{length})^{2},\quad\textrm{dim}(s_{Rd}^{(c)})=\textrm{force}\times(\textrm{length})^{-2}.
\]
\subsection{`Discrete' parameters}\label{62}
On resuming the spring-and-stick models of an A- and a Z- CNT assembled in Section 3, we set:
\begin{equation}\label{rigtrax}
s^{(d)}_{Ax}:=\frac{F}{\Delta H^M/H^M}\quad\textrm{and}\quad \nu^{(d)}_{Ax}:=-\frac{\Delta P^M/P^M}{\Delta H^M/H^M}\,,
\end{equation}
for the discrete axial stiffness and axial contraction modulus, where $H^M,P^M$ and $\Delta H^M,\Delta P^M$ are the undeformed height and perimeter of a module and their changes under the applied loads. While the first of these definitions parallels the first of \eqref{rigtrap} in a self-explanatory manner, the second one requires a motivation: ours is that, for $n$ sufficiently large and axisymmetric loadings,
\[
\Delta P^M/P^M\approx \Delta \rho_0/\rho_0\,.
\]
For an explicit evaluation of $s^{(d)}_{Ax}$ and $ \nu^{(d)}_{Ax}$, we proceed case by case. We find (Appendix, Subsection \ref{4}):
\begin{itemize}
\item (armchair case)
\begin{equation}\label{AsA}
s^{(d,A)}_{Ax}=\frac{4n}{\sqrt 3+\frac{1}{4\sqrt
3}\left(1+\frac{3}{2\tan^2\beta} \right)^{-1}\frac{k_a
s^2}{k_s}}\,k_a s
\end{equation}
and
\begin{equation}\label{AnuA}
\quad\nu^{(d,A)}_{Ax}=-\frac{1-\frac{1}{4}\left(1+\frac{3}{2\tan^2\beta} \right)^{-1}\frac{k_a s^2}{k_s}}{3+\frac{1}{4}\left(1+\frac{3}{2\tan^2\beta} \right)^{-1}\frac{k_a s^2}{k_s}}
\end{equation}
\item (zigzag case)
\begin{equation}\label{ZsA}
s^{(d,Z)}_{Ax}=\frac{n}{\frac{3}{4}+\frac{1}{8\left(1+\frac{3}{2}\frac{\cos^4(\gamma/2)}{\sin^2\beta}\right)}\,\frac{k_a}{k_s}s^2}\,k_as
\end{equation}
and
\begin{equation}\label{ZnuA}
\nu^{(d,Z)}_{Ax}=-\frac{\frac{1}{4}+\frac{\sqrt{3}(1+\cos\beta)}{8(5+\cos\beta)}
\frac{k_a
s^2}{k_s}}{\frac{3}{4}+\frac{1}{8\left(1+\frac{3}{2}\frac{\cos^4(\gamma/2)}{\sin^2\beta}\right)}\frac{k_a
s^2}{k_s}}\,.
\end{equation}
\end{itemize}

An alternative way to evaluate discrete strains and stiffnesses
is to make use of Lam\'e's theorem of work and energy (\cite{Gu}, Section 28). As is well known, this theorems guarantees that, in a linearly elastic equilibrium problem,  \emph{the load work is twice the stored energy}. Accordingly, for axial traction,
\[
F\Delta H^M=2\,U^M_{Ax},\quad\textrm{with}\;\, U^M_{Ax}=\textrm{energy stored in a module}.
\]
Hence, on applying definition  $\eqref{rigtrax}_1$, we find:
\[
s^{(d)}_{Ax}=\frac{F^2H^M}{2\, U^M_{Ax}}\,;
\]
in particular, when $U^M$ is evaluated with the use of \eqref{armen} and \eqref{moden}, \eqref{AsA} follows.

To estimate torsion
stiffness, we invoke Lam\'e's theorem and write, for a
module,
$$
T\Theta H^{M}=2U^{M}_{To},
$$
where $T$ is the applied torque and $\Theta$ is the twist angle per unit of length; on inserting this relation in \eqref{rigtors}, we find:
$$
s_{To}^{(d)}=\frac{T^2H^{M}}{2\,U^{M}_{To}}.
$$
For an ACNT, $T=2n\,h\rho_0$, where $h$ has been defined in Section \ref{torz} (see Fig. \ref{tors} and equations \eqref{fortor}). On using \eqref{rozzo}, \eqref{entor}, \eqref{soluz} and \eqref{soluz2}, we find:
\begin{equation}\label{sto}
s_{To}^{(d,A)}=\frac{9(\sqrt{3}/2\pi^2)n^3}{\cos(\gamma-\vartheta)+2\frac{\sin^2(\gamma-\vartheta)}{\sin\gamma}+2\left(2\frac{\sin^2\beta\sin^2(\gamma-\vartheta)}{\sin^2\gamma}+3\left(\frac{1}{2}\cos\vartheta-\frac{\rho_0}{s}\frac{\sin^2\vartheta}{\sin\gamma}\right)^2
\right)\frac{k_a s^2}{k_s}}\,k_a s^3\,.
\end{equation}

In the case of radial applied forces, the Lam\'e theorem yields:
\begin{equation}\label{lamet}
N^{M} p\,w_{Rd}=2\, U^{M}_{Rd},
\end{equation}
where $w_{Rd}=$ radial displacement. It is the matter of a simple computation (Appendix, Subsection \ref{62}) to arrive at the following
expressions for the discrete radial stiffnesses of a A- and Z-
CNT:
\begin{equation}\label{dradsti}
s_{Rd}^{(d, A)}=
\frac{32\frac{\sin^2\gamma}{\sin^2\varphi}\,n}{\sqrt{3}\pi\left(5+4\cos2\gamma+\frac{6}{8+3\frac{\cos^2\gamma}{\sin^2\beta}}\frac{k_as^2}{k_s}
\right)}\,k_a s^{-1},
\end{equation}
\begin{equation}\label{dradstiz}
s_{Rd}^{(d,
Z)}=\frac{16n(17-8\cos2\beta+12\cos\gamma+3\cos2\gamma)\frac{\sin^2\frac{\gamma}{2}}{\sin^2\beta}}{3\pi\left(6+9\frac{\cos^4\frac{\gamma}{2}}{\sin^2\beta}+\frac{k_as^2}{k_s}
\right)}\,k_a s^{-1};
\end{equation}
likewise, it is not difficult to show that
\begin{equation}\label{dcontr}
\nu_{Rd}^{(d,A)}=\frac{3}{\sin^2\beta}\,\frac{2\,\frac{k_as^2}{k_s}-11+2\cos2\beta(4-\frac{k_as^2}{k_s})-3\cos2\gamma}{4(3\frac{k_as^2}{k_s}+4)+2\cos\gamma\left(
64+3\cos\gamma\frac{1+8\cos\gamma}{\sin^2\beta}\right)}\,,
\end{equation}
and
\begin{equation}\label{dcontrz}
\nu_{Rd}^{(d,Z)}=-\frac{1}{3}+\frac{4}{3\left(1+\frac{k_s}{k_as^2}\left(6+9\frac{\cos^4\frac{\gamma}{2}}{\sin^2\beta}
\right)\right)}\,.
\end{equation}
\subsection{Elastic moduli, thickness, and radius, of CNT-like shells}\label{costpar}
\begin{figure}[h]
\centering
\includegraphics[scale=0.8]{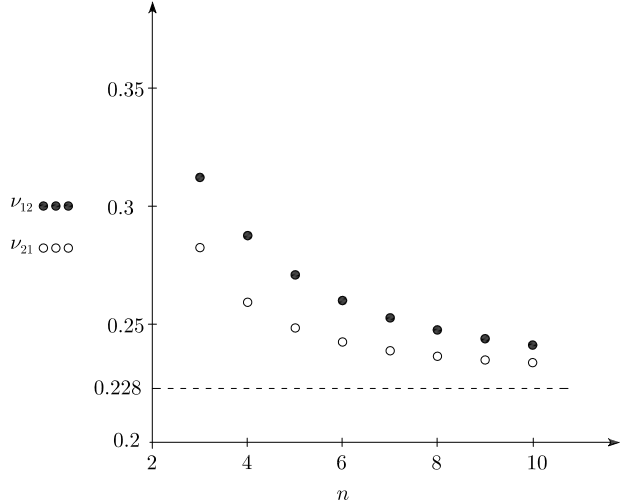}
\caption{The contraction moduli $\nu_{21}\;(\circ)$ and $\nu_{12}\;(\bullet)$ of ACNT-like shells.} \label{poisson}
\end{figure}

i. \underline{Contraction moduli}. We choose the cylinder axis of a CNT-like shell parallel to the  roll-up axis of an ACNT, and  require that
\begin{equation}\label{pure}
\nu_{Rd}^{(c)}=\nu_{Rd}^{(d,A)},
\end{equation}
whence,  by $\eqref{radsti}_2$ and \eqref{dcontr}, we obtain that
\begin{equation}\label{nu21}
\nu_{21}=\frac{3}{\sin^2\beta}\,\frac{2\,\frac{k_as^2}{k_s}-11+2\cos2\beta\,(4-\frac{k_as^2}{k_s})-3\cos2\gamma}{4\left(3\frac{k_as^2}{k_s}+4\right)+2\cos\gamma\left(
64+3\cos\gamma\frac{1+8\cos\gamma}{\sin^2\beta}\right)}\,.
\end{equation}
Likewise, on choosing the cylinder axis of a CNT-like shell made of the same orthotropic material parallel to the roll-up axis of a ZCNT, we get:
\begin{equation}\label{nu12}
\nu_{12}=\nu_{Rd}^{(d,Z)}=-\frac{1}{3}+\frac{4}{3\left(1+\frac{k_s}{k_as^2}\left(6+9\frac{\cos^4\frac{\gamma}{2}}{\sin^2\beta}
\right)\right)}\,.
\end{equation}
The dependence of these two Poisson-like moduli on the size parameter $n$ is shown in Figure \ref{poisson};
note that
\begin{equation}\label{nuinf}
\lim_{n\rightarrow \infty}\nu_{21}=\lim_{n\rightarrow \infty}\nu_{12}=:\nu^\infty=0.228.
\end{equation}

\noindent ii. \underline{Radius}. With a view toward determining the \emph{effective radii} $\rho_o^A$ and $\rho_o^Z$ of A- and Z- CNT-like shells, we observe that,  by the use of $\eqref{rigtra}_1$, $\eqref{radsti}_1$ and $\eqref{rigtra}_2$, and for \eqref{ennu} to hold, we find:
$$
\frac{s_{Ax}^{(c)}}{s_{Rd}^{(c)}}=2\pi\rho_o^2\frac{\delta(\eps/\rho_o)}{1-\nu_{12}\nu_{21}(1-\delta(\eps/\rho_o)}\frac{E_1}{E_2}=2\pi\rho_o^2\frac{\nu^{(c)}_{Ax}}{\nu_{12}}\frac{E_1}{E_2}=2\pi\rho_o^2\frac{\nu^{(c)}_{Ax}}{\nu_{21}},
$$
whence, thanks to $\eqref{radsti}_2$, we arrive at
$$
\frac{s_{Ax}^{(c)}}{s_{Rd}^{(c)}}=2\pi\rho_o^2\frac{\nu_{Ax}^{(c)}}{\nu_{Rd}^{(c)}}\quad\Leftrightarrow\quad
\rho_o=\sqrt{\frac{1}{2\pi}\frac{s_{Ax}^{(c)}}{s_{Rd}^{(c)}}\frac{\nu_{Rd}^{(c)}}{\nu_{Ax}^{(c)}}}\,.
$$
We then set:
\begin{equation}\label{radius}
\rho_o^A:=\sqrt{\frac{1}{2\pi}\frac{s_{Ax}^{(d,A)}}{s_{Rd}^{(d,A)}}\frac{\nu_{Rd}^{(d,A)}}{\nu_{Ax}^{(d,A)}}},\quad\rho_o^Z:=\sqrt{\frac{1}{2\pi}\frac{s_{Ax}^{(d,Z)}}{s_{Rd}^{(d,Z)}}\frac{\nu_{Rd}^{(d,Z)}}{\nu_{Ax}^{(d,Z)}}}\,.
\end{equation}
Both effective radii depend on the size parameter $n$ in a complex manner. But, as Fig. \ref{raggio} shows,
\begin{figure}[h]
\centering
\includegraphics[scale=0.8]{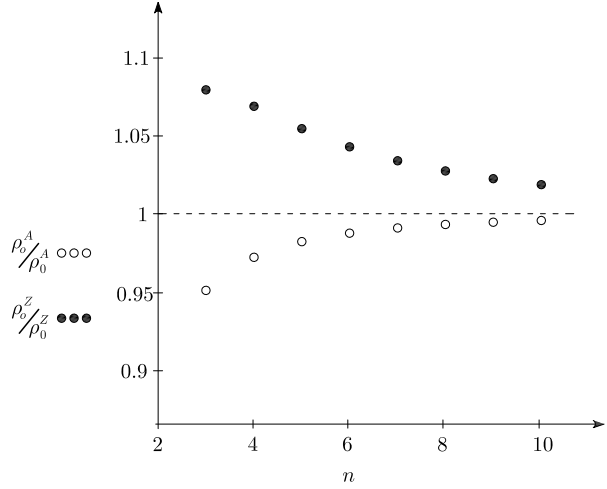}
\caption{Ratio of effective and geometrically necessary radii of A-$(\circ)$ and Z-$(\bullet)$ CNT-like shells.}
\label{raggio}
\end{figure}
neither of them differs much, whatever $n$, from the corresponding value of the geometrically necessary radius; moreover, as $n$ grows bigger they both tend to it:
\[
\lim_{n\rightarrow \infty}\rho_o^A/\rho_0^A=\lim_{n\rightarrow \infty}\rho_o^Z/\rho_0^Z=1.
\]

\noindent iii. \underline{Thickness}. We begin by laying down the following equality:
\begin{equation}\label{sp3}
\frac{\delta(\varepsilon^A/\rho_o^A)}{1-\nu_{12}\nu_{21}\left(1-
 \delta(\varepsilon^A/\rho_o^A)\right)}\,\nu_{12}=\nu_{Ax}^{(d,A)},
\end{equation}
where the values for $\nu_{21},\nu_{12}$ and $\rho_o^A$ are chosen as agreed in i. and ii.,  and give it the role of an implicit equation that we solve for the unknown $\eps^{A}$.\footnote{With the use of \eqref{deltaeps}, it is the matter of a short computation to give \eqref{sp3} the following form:
\[
\frac{1}{2x}\log\frac{1+x}{1-x}=\frac{\nu_{21}}{\nu_{Ax}^{(d,A)}}\,;
\]
for $\bar x$ the unique positive solution of this equation, we have that $\varepsilon^A=\rho_o^A\,\bar x$.
}

\begin{figure}[h]
\centering
\includegraphics[scale=0.8]{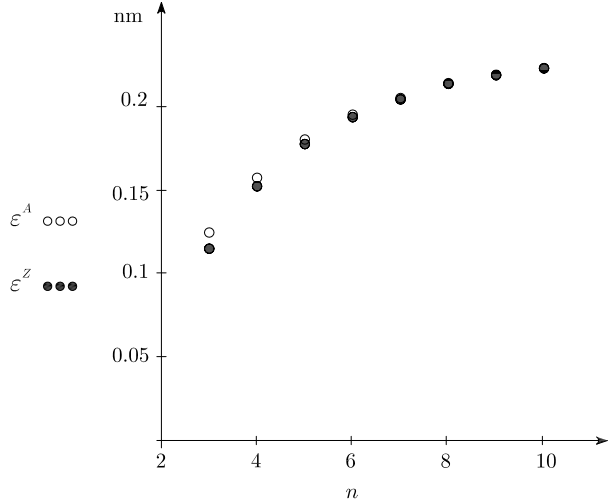}
\caption{The effective thickness of A-$(\circ)$ and Z-$(\bullet)$ CNT-like shells.}
\label{epsilon}
\end{figure}

As to $\varepsilon^Z$, we proceed as follows. Firstly, we equate the continuous and discrete radial stiffnesses of both A- and Z- CNTs:
\begin{equation}\label{radsp}
\begin{aligned}
&2E_2\varepsilon^A=s_{Rd}^{(d,A)}\rho_o^A\delta(\varepsilon^A/\rho_o^A),\\
&2E_1\varepsilon^Z=s_{Rd}^{(d,Z)}\rho_o^Z\delta(\varepsilon^Z/\rho_o^Z)
\end{aligned}
\end{equation}
(cf. $\eqref{radsti}_1$, \eqref{dradsti}, and \eqref{dradstiz}), whence, with the use of \eqref{ennu}, we arrive at
$$
\frac{\varepsilon^Z}{\varepsilon^A}=\frac{\nu_{21}}{\nu_{12}}\frac{s_{Rd}^{(d,Z)}}{s_{Rd}^{(d,A)}}\frac{\rho_o^Z}{\rho_o^A}
\frac{\delta(\varepsilon^Z/\rho_o^Z)}{\delta(\varepsilon^A/\rho_o^A)}.
$$
Finally, given that \eqref{sp3} implies that
\begin{equation}\label{deltaA}
\delta(\varepsilon^A/\rho_o^A)=\frac{\nu_{Ax}^{(d,A)}}{\nu_{12}}\frac{1-\nu_{12}\nu_{21}}{1-\nu_{21}\nu_{Ax}^{(d,A)}},
\end{equation}
and that a completely similar relation holds for $\delta(\varepsilon^Z/\rho_o^Z)$, we obtain:
\begin{equation}\label{spessZ}
\varepsilon^Z=\varepsilon^A\,\frac{s_{Rd}^{(d,Z)}}{s_{Rd}^{(d,A)}}
\frac{\nu_{Ax}^{(d,Z)}}{\nu_{Ax}^{(d,Z)}}\frac{1-\nu_{21}\nu_{Ax}^{(d,A)}}{1-\nu_{12}\nu_{Ax}^{(d,Z)}}\frac{\rho_o^Z}{\rho_o^A}\,.
\end{equation}
Perusal of \eqref{sp3} and \eqref{spessZ} yields for the effective thicknesses $\varepsilon^A$ and $\varepsilon^Z$ the dependences on the chirality index $n$ shown in fig. \ref{epsilon}.

\begin{figure}[h]
\centering
\includegraphics[scale=0.8]{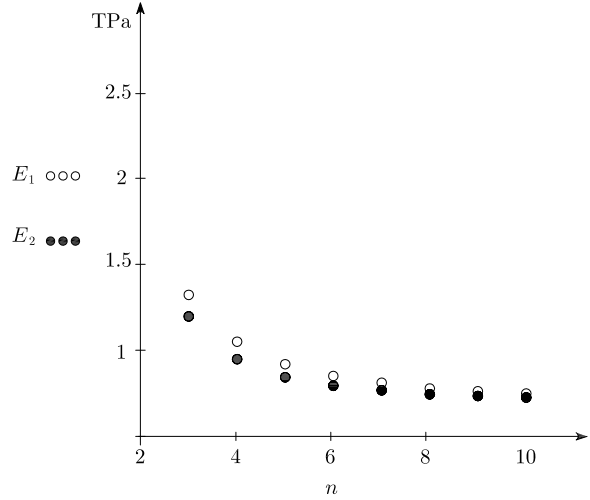}
\caption{The extension moduli of CNT-like shells.} \label{young}
\end{figure}

\begin{figure}[h]
\centering
\includegraphics[scale=0.8]{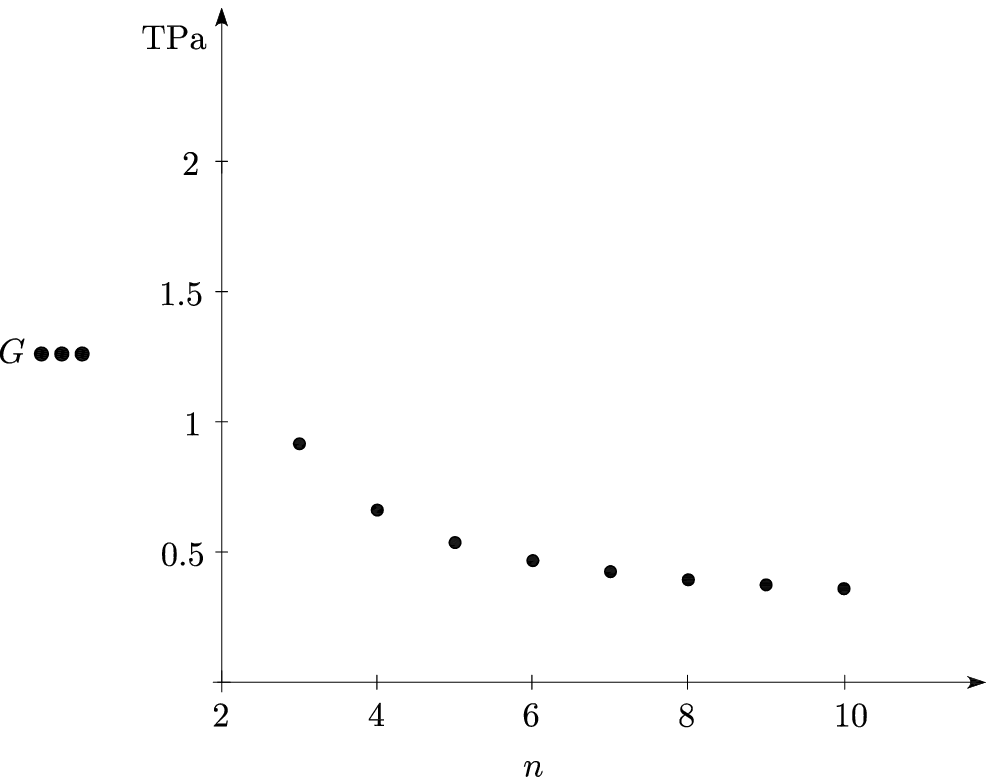}
\caption{The shear modulus of ACNT-like shells.} \label{shear}
\end{figure}

\vskip 6pt
\noindent iv. \underline{Extension moduli}. Once effective radii and thicknesses have been computed, one way (there are others) to find the constitutive moduli $E_1,E_2$ is to combine \eqref{rigtra} and \eqref{deltaA}, under the assumption that $s^{(c,A)}_{Ax}=s^{(d,A)}_{Ax}$ and
$\nu^{(c,A)}_{Ax}=\nu^{(d,A)}_{Ax}$,  so as to obtain:
\begin{equation}
E_1(2\varepsilon^A)=\frac{s_{Ax}^{(d,A)}}{2\pi\rho_o^A}\frac{1-\nu_{12}\nu_{21}}{1-\nu_{21}\nu_{Ax}^{(d,A)}}\,;
\end{equation}
a germane relation holds for $E_2$, namely,
\begin{equation}
E_2(2\varepsilon^Z)=\frac{s_{Ax}^{(d,Z)}}{2\pi\rho_o^Z}\frac{1-\nu_{12}\nu_{21}}{1-\nu_{21}\nu_{Ax}^{(d,Z)}}\,.
\end{equation}
The dependence of both $E_1(2\varepsilon^A)$ and $E_2(2\varepsilon^Z)$ on $n$ is shown in Figure \ref{ethic}; Figure \ref{young}
shows how the values of both $E_1$ and $E_2$ decay with $n$, their difference becoming smaller and smaller.

\noindent v. \underline{Shear modulus}.
By equating the
continuum and discrete evaluations \eqref{rigtors} and \eqref{sto} of the torsional stiffness, we find:
\begin{equation}\label{G}
G=\frac{s_{To}^{(d)}}{J(\eps)}=\frac{s_{To}^{(d)}}{4\pi(\rho_o^A)^3\varepsilon^A}
\left(1+(\varepsilon^A/\rho_o^A)^2\right)
\end{equation}
(see Figure \ref{shear}).

\noindent vi. \underline{Consistency checks}. The correctness of our computations can be checked in a number of ways.
For one, \eqref{ennu}, a key feature of our description of the constitutive response of orthotropic shell, is satisfied exactly, to all practical puposes, by the quadruplet $(E_1,E_2,\nu_{12},\nu_{21})$ computed in the preceding subsections by equating continuous and discrete stiffnesses and contraction moduli. In fact, by an appropriate re-use of the computations made to construct Figures \ref{poisson} and \ref{young}, one can show that, whatever the index $n$, $E_1/E_2-\nu_{12}/\nu_{21}\approx 5\times 10^{-16}$.

For two, the thickness ratio $\varepsilon^Z/\varepsilon^A$ can be evaluated in a way alternative to the one implicit in \eqref{epsilon},  namely,
\begin{equation}\label{altern}
\frac{\varepsilon^Z}{\varepsilon^A}=2\pi\rho_o^A\rho_o^Z\frac{s_{Rd}^{(d,Z)}}{s_{Ax}^{(d,A)}}
\frac{\nu_{Ax}^{(d,Z)}}{\nu_{21}}\frac{1-\nu_{21}\nu_{Ax}^{(d,A)}}{1-\nu_{12}\nu_{Ax}^{(d,Z)}},\footnote{\eqref{altern} is arrived at without presuming that  \eqref{ennu} holds.
}
\end{equation}
with numerically identical results. Interestingly, no matter whether \eqref{epsilon} or \eqref{altern} is used, one has that
\[
\lim_{n\rightarrow\infty} \frac{\varepsilon^Z}{\varepsilon^A}=1,
\]
that is to say, the thickness of A- and Z- CNTs of larger and larger radius tends to be the same.

\section{Conclusions}\label{concl}
We have presented a procedure to determine  the constitutive moduli and the geometric parameters to be associated with a single-wall carbon nanotube of given chirality in order to model it as a linearly elastic orthotropic cylindrical shell; our findings are summarized in the following table, for various values of the chirality index $n$, where we have set $h^A:=2\,\varepsilon^A,\, h^Z:=2\,\varepsilon^Z$.
\vskip 20pt
{\begin{center}
\centering
{\begin{tabular}{|c|c|c|c|}
\hline $n$ & 6 & 10 & 20 \\
\hline $E_1$ (TPa) & 0.832 & 0.736 & 0.709 \\
\hline $E_2$ (TPa)& 0.784 & 0.729 & 0.706 \\
\hline $\nu_{12}$  & 0.260 & 0.241 & 0.232 \\
\hline $\nu_{21}$ & 0.242 & 0.233 & 0.230  \\
\hline $G$ (TPa) & 0.424 & 0.359 & 0.307  \\
\hline $\h^A$ (nm)& 0.390 & 0.446 & 0.466 \\
\hline $\rho_o^A$ (nm) & 0.402 & 0.675 & 1.354 \\
\hline
\end{tabular}
\hspace{0.5cm}
\begin{tabular}{|c|c|c|c|}
\hline $n$ & 6 & 10 & 20 \\
\hline $E_1$ (TPa) & 0.784 & 0.729 & 0.706 \\
\hline $E_2$ (TPa)& 0.832 & 0.736 & 0.709 \\
\hline $\nu_{12}$  & 0.242 & 0.233 & 0.230 \\
\hline $\nu_{21}$ & 0.260 & 0.241 & 0.232  \\
\hline $G$ (TPa) & 0.424 & 0.359 & 0.307  \\
\hline $h^Z$ (nm) & 0.388 & 0.446 & 0.466 \\
\hline $\rho_o^Z$ (nm) & 0.245 & 0.399 & 0.787 \\
\hline
\end{tabular}
}
\end{center}
\vskip 10pt
\centerline{Table 1. Elastic moduli, thickness and radius of CNT-like shells (A-, left; Z-, right).}
\vskip 10pt

As anticipated in our introductory section, the list of constitutive and geometric parameters appropriate for a SWCNT of arbitrary chirality pair $(n,m)$ can be deduced for either one of the two lists in Table 1. A feature of our procedure is that each of these parameters, in addition to chirality, depends only on the two material constants modulating the harmonic approximations of the nanoscopic bonding energies and on the length of a C-C bonds.\footnote{These  three nanoscopic bits of information can be put together under form of a dimensionless characteristic number, namely, $\frac{k_a s^2}{k_s}=13.101$, measuring the relative importance of stretching and angle-changing energies.}

Here is an interesting conclusion that can be drawn from our results in the so-called `graphene limit', that is, for index $n\rightarrow \infty$. We have that
\[
E_1/E_2=\frac{\rho_o^{A}}{\rho_o^{Z}}\frac{s_{Ax}^{(d,Z)}}{s_{Ax}^{(d,A)}}\frac{\varepsilon^A}{\varepsilon^Z}\frac{1-\nu_{12}\nu_{21}(1-\delta(\varepsilon^A/\rho_o^A))}{1-\nu_{12}\nu_{21}(1-\delta(\varepsilon^Z/\rho_o^Z))}\;\rightarrow 1,\;\;\, \nu_{12}/\nu_{21}\;\rightarrow 1,\;\;\, h^A/h^Z\;\rightarrow 1;
\]
moreover,
\[
\begin{aligned}
\lim_{n\rightarrow\infty}Gh^A=:G^\infty h^\infty&=0.143\,{\rm TPa}\times{\rm nm},\\
\lim_{n\rightarrow\infty}E_1h^A=: E^\infty h^\infty&=0.332\,{\rm TPa}\times{\rm nm},\\
\nu^\infty&=0.228
\end{aligned}
\]
(recall \eqref{nuinf}). Now, it is to check that
\[
\frac{E^\infty h^\infty}{2(1+\nu^\infty)}=0.135\,{\rm TPa}\times{\rm nm}\neq G^\infty h^\infty,
\]
whereas equality should hold when graphene, modeled within linear plane elasticity, is regarded as isotropic.

Another feature of our study is that it allows for determining in a mutually consistent manner the values of both wall thickness and Young modulus  of a SWCNT. What value to take for thickness has been an issue in the nanotube community for long. Needless to say, an evaluation of effective thickness is necessary to apply any shell theory, ours being no exception.
In the following table, a subset of Table 1 of \cite{HWH}, values for wall thickness and Young
modulus of a SWCNT are listed, as determined by a number of different procedures; we have added a last column, for the values of the \emph{tensile  rigidity} \cite{HWH}, that is, the product Young's modulus$\,\times\,$wall thickness.
\vskip 20pt
\scriptsize{
\hspace{-0.9cm}\begin{tabular}{|c|c|c|c|c|}
  \hline
   &  & Wall  & Young's  &  Tensile  \\
   Authors & Method  & thickness $h$  &  modulus $E$  &  rigidity $Eh$  \\
  &  &  (nm) &  (TPa) &  (TPa$\times$nm)   \\
  \hline
  Lu \cite{Lu}& Molecular dynamics & 0.34 & 0.974 & 0.331 \\
 Hernandez et al. \cite{He}& Tight-binding molecular dynamics & 0.34 & 1.24 & 0.422 \\
  Li and Chou \cite{Li}& Structural mechanics: stiffness & 0.34 & 1.01 & 0.343 \\
   & matrix method &  &  &  \\
 Jin and Yuan \cite{JY}& Molecular dynamics & 0.34 & 1.238 & 0.421 \\
    Yakobson \cite{Y} et al. & Molecular dynamics & 0.066 & 5.5 & 0.363 \\
  Zhou et al. \cite{ZZ} & Tight-binding model & 0.074 & 5.1 & 0.377 \\
  Kudin et al. \cite{KSY} & \emph{Ab initio} computations & 0.089 & 3.859 & 0.343 \\
  Tu and Ou-Yang \cite{TO}& Local density & 0.075 & 4.7 & 0.352 \\
   & approximation model &  &  &  \\
 Vodenitcharova and & Ring theory continuum mechanics & 0.0617 & 4.88 & 0.301 \\
  Zhang \cite{VZ} &  &  &  &  \\
  Pantano et al. \cite{PPB} & Continuum shell modeling & 0.075 & 4.84 & 0.363 \\
  Wang et al. \cite{WZ} & \emph{Ab initio} calculation & 0.0665 & 5.07 & 0.337 \\
  \hline
\end{tabular}}
\normalsize
\vspace{.5cm}

\noindent Interestingly, while the reported values for wall thickness and Young modulus display a large scattering -- the so-called
\emph{Yakobson's paradox} \cite{Sh} --
the tensile rigidity has a very small variance, irrespectively of the method used to evaluate the factors. We regard this evidence as an implicit resolution of the paradox, in that it makes evident what one can sensibly try and evaluate: the stiffness parameters defined in Section \ref{effpar}, or others of the same sort, in two ways to be juxtaposed, discrete and continuous.

Note that the values furnished by our theory for the tensile rigidity comply with those in the table whatever the chirality index; Figure \ref{ethic} 
summarizes our findings.
\begin{figure}[h]
\centering
\includegraphics[scale=0.8]{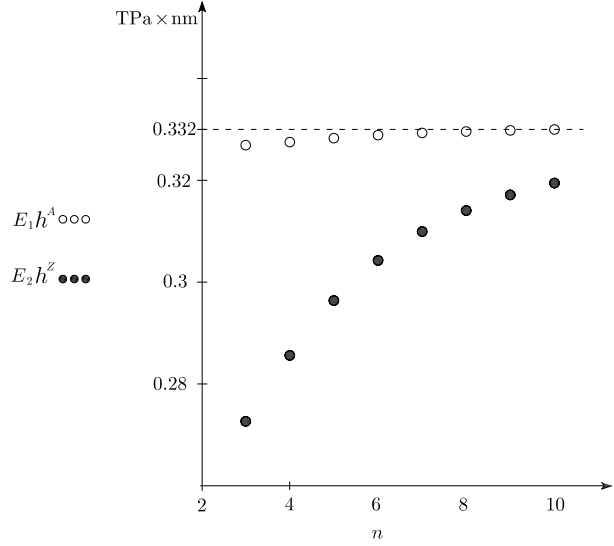}
\caption{The tensile rigidity of A-$(\circ)$ and Z-$(\bullet)$ CNT-like shells.} \label{ethic}
\end{figure}
%

Finally, it is worth noticing that, in the light of our present theory, two
notions of thinness can be given for a SWCNT. The first, \emph{geometric thinness}, is expressed by formula \eqref{thin}; it depends exclusively on the chirality index $n$, by way of the ratio of the length $s$ of the C-C bond and the geometrically necessary radius $\rho_0$. The second, \emph{effective thinness}, is $\eps/\rho_o$; here both
$\eps$ and $\rho_o$ depend in a complicated way, in addition to $s$ and $n$, on
the constitutive nanoscopic stiffness parameters $k_a$ and $k_s$. Figure \ref{thinness} shows that the effective thinness, while being greater than the geometric one, approaches zero as $n$ grows in the same $n^{-1}-$ way as the latter; it also shows that for small radii CNT-like shells are far from being thin.
\begin{figure}[h]
\centering
\includegraphics[scale=0.8]{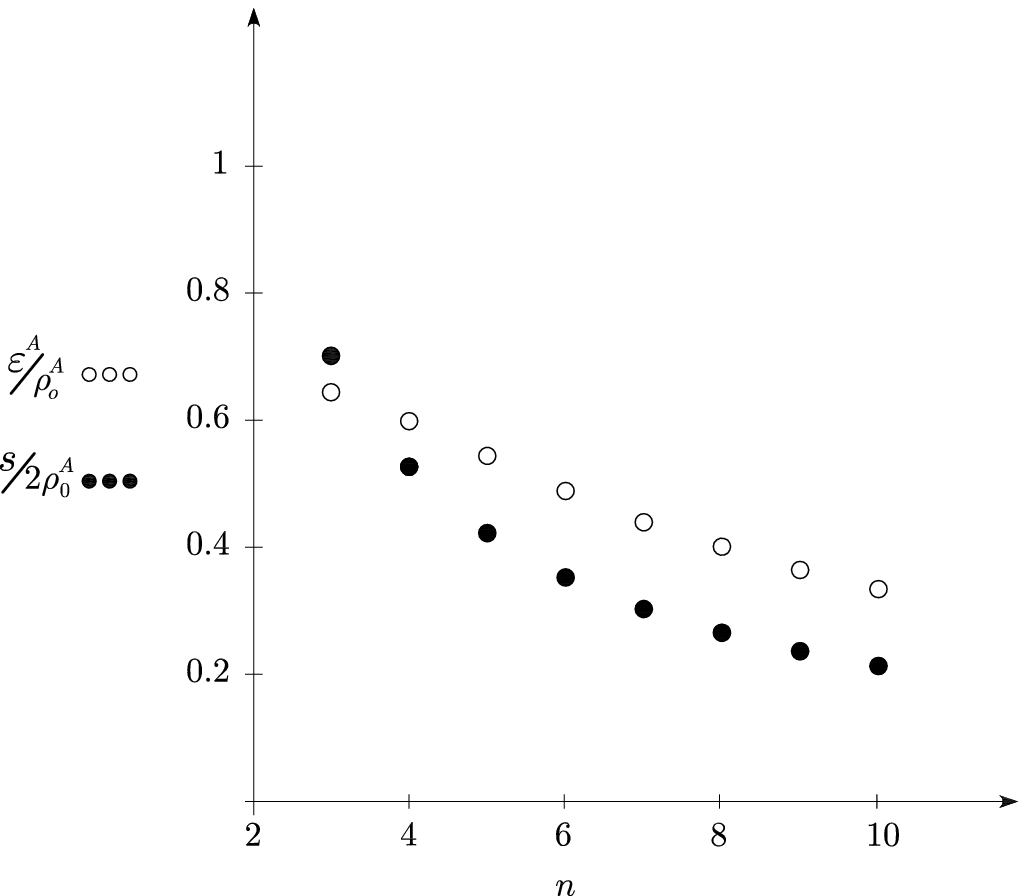}
\caption{Geometric $(\bullet)$ and effective $(\circ)$ thinness.} \label{thinness}
\end{figure}

\scriptsize{
\section{Appendix. Ancillary Computations}
\subsection{Computations relative to Section \ref{ata}}\label{app1}
Note that
\[
|\bb\times\cb|=\sin\beta=|\cb\times\eb|,\quad |\cb\times\db|=\sin\alpha,
\]
two relations that will be useful to derive \eqref{eulmod} here below.
Combination with $\eqref{forze}_1$, \eqref{coup1}-\eqref{angvar}, and \eqref{forba2}, allows us to give \eqref{eul} the following form:
\begin{equation}\label{eulmod}
\cb\times\big(s\,\fb_C+2\tau_\alpha\,\db+\tau_\beta\,(\eb-\bb)\big)=\mathbf{0};
\end{equation}
thus, this equation has only two scalar counterparts. Now, with the use of $\eqref{posiz}_2$, \eqref{forbal}, \eqref{forze}, and  \eqref{forba2}, one obtains:
\begin{equation}\label{equil2}
\begin{aligned}
&\cb\times\fb_C=(\fb_C\cdot\cb_3)\cos\frac{\alpha}{2}\sin\gamma\cb_1+\left((\fb_C\cdot\cb_3)\cos\frac{\alpha}{2}\cos\gamma-\frac{1}{2}(\fb_B\cdot\cb_1)\sin\frac{\alpha}{2}
\right)\cb_2+\\
&\hspace{2.3cm}
+\frac{1}{2}(\fb_B\cdot\cb_1)\cos\frac{\alpha}{2}\sin\gamma\cb_3;
\end{aligned}
\end{equation}
moreover, it follows from \eqref{posiz} and \eqref{evect} that
\begin{equation}
\begin{aligned}
&\cb\times\db=-\sin\alpha(\sin\gamma\,\cb_1+\cos\gamma\,\cb_2),\\
&\cb\times\eb=-\sin\frac{\alpha}{2}(\sin2\gamma\,\cb_1+\cos2\gamma\,\cb_2)-\cos\frac{\alpha}{2}\sin\gamma\,\cb_3,\\
&\cb\times\bb=\sin\frac{\alpha}{2}\,\cb_2-\cos\frac{\alpha}{2}\sin\gamma\,\cb_3.
\end{aligned}
\end{equation}
Consequently, in view also of \eqref{coup1}, \eqref{coup2} and \eqref{angvar}, the vector equation \eqref{eulmod} yields the
following scalar system:
\begin{equation}\label{ACeq}
\left\{\begin{array}{c}\fb_B\cdot\cb_1=0,\qquad\qquad \qquad\qquad\qquad\qquad\qquad\qquad\\\displaystyle{s\cos\frac{\alpha}{2}(\fb_C\cdot\cb_3)-2k_s\left(\Delta\alpha-\Delta\beta\,\frac{\sin\frac{\alpha}{2}}{\sin\beta}\cos\gamma\right)=0.}\end{array}\right.
\end{equation}
Thus, as to applied forces, we have:
\begin{equation}\label{applforz}
\fb_B=\mathbf{0},\quad \fb_C=-\fb_D=f\cb_3,
\end{equation}
These developments lead to \eqref{allung}; \eqref{deltavarib} is arrived at by combining the second of \eqref{ACeq} with
$\eqref{cos}_1$ and \eqref{deltavari} and \eqref{applforz}.

\subsection{Computations relative to Section \ref{atz}}\label{Aatz}
On taking
\eqref{zigpos} into account, one finds that
$$
\begin{aligned}
s^{-1}\pb_B\times\fb_B&=-(\fb_{B}\cdot\cb_2)\cos\alpha\cb_1+\left((\fb_{B}\cdot\cb_1)l\cos\alpha+\frac{1}{2}(\fb_{C}\cdot\cb_3)\sin\alpha\right)\cb_2\\&+(\fb_{B}\cdot\cb_2)\sin\alpha\cb_3;
\end{aligned}
$$
moreover,
$$
\begin{aligned}
&\bb\times\db=-\sin\alpha\cos\alpha\sin\gamma\cb_1-\sin\alpha\cos\alpha(1+\cos\gamma)\cb_2+\sin^2\alpha\sin\gamma\cb_3,\\
&\bb\times\mb=-\sin\alpha\cos\alpha\sin\gamma\cb_1-(1+\cos\gamma)\sin\alpha\sin\gamma\cb_2+\sin^2\alpha\sin\gamma\cb_3,
\\
&\bb\times\cb_3=-\sin\alpha\cb_2.
\end{aligned}
$$
With the use of these relations, equation \eqref{Zeq} yields the following scalar system:
\begin{equation}\label{sis2}
\left\{\begin{array}{c}\fb_{B}\cdot\cb_2=0,\qquad\qquad\qquad\qquad\qquad\qquad\qquad\qquad\qquad\qquad\qquad\qquad\qquad\qquad
\\
\displaystyle{(\fb_{B}\cdot\cb_1)s\cos\alpha+\frac{1}{2}(\fb_{C}\cdot\cb_3)s\sin\alpha
-2k_s\frac{\Delta\beta}{\sin\beta}(1+\cos\gamma)\sin\alpha\cos\alpha-2k_s\Delta\alpha=0.}\end{array}\right.
\end{equation}
Now, the first of \eqref{sis2} and the last of \eqref{symzig} imply that  $\fb_{D}\cdot\ab\times\cb_3=0$; with this and the penultimate of \eqref{symzig}, the force balance \eqref{forbal} imply that  $\fb_{B}\cdot\cb_1=0$. Summing up, the force system is:
\[
\fb_B=\fb_D=-\fb_C=-f\cb_3,
\]
whence the result in \eqref{allung_zif}. Moreover, $\eqref{sis2}_2$ reduces to
\[
\frac{1}{2}\frac{fs}{k_s}\,\sin\alpha
-2\frac{\Delta\beta}{\sin\beta}(1+\cos\gamma)\sin\alpha\cos\alpha-2\Delta\alpha=0,
\]
whence, with the use of \eqref{Zdiff}, we obtain \eqref{deltavarib_zif}.

\subsection{Computations relative to Section \ref{torz}}\label{Atorz}
\noindent{\textbf{(i) \emph{Preparatory developments.}}}
With a view toward formulating the  equilibrium conditions for the $AC$ and $AD$ sticks, we supplement \eqref{coptorter} with the following further guess:
\begin{equation}\label{copptor}
\Delta\beta^{u}(A)=-\Delta\beta^{d}(A),
\end{equation}
where $\Delta\beta^u$ ($\Delta\beta^d$) denotes the change in angle between the $AC$ ($AD$) and $AB$ sticks.
Moreover, this time in preparation for formulating the first moment balance,  we pick the pole $P$ shown in Figure \ref{tors} and, with the help of Figure \ref{tors1}, we specify as follows the relative position vectors of the points where forces are applied:
\begin{equation}\label{altrepos}
\begin{aligned}
\pb_C&=\hb+\rb_C,\quad\pb_D=-\hb+\rb_C,\\
\pb_E&=\hb+\rb_E,\quad\pb_F=-\hb+\rb_E,
\end{aligned}
\end{equation}
 with
 \begin{equation}\label{vectors}
 \begin{aligned}
 \hb&=s\sin\frac{\alpha}{2}\cb_3,\\
 \rb_C&=-\rho_0(\sin\vartheta\cb_1+\cos\vartheta\cb_2)=\rb_A+s\cos\frac{\alpha}{2}\ab,\\ \rb_E&=\rho_0(\sin\vartheta\cb_1-\cos\vartheta\cb_2)= \rb_B+s\cos\frac{\alpha}{2}\ab^\prime,\\
 \rb_A&=-\rho_0(\cos\varphi\cb_1+\sin\varphi\cb_2),\quad \rb_B=\rho_0(\cos\varphi\cb_1-\sin\varphi\cb_2),\\ \ab&=-\cos\gamma\cb_1+\sin\gamma\cb_2,\quad \ab^\prime=\cos\gamma\cb_1+\sin\gamma\cb_2,
 \end{aligned}
 \end{equation}
and
\[
\rho_0=\frac{3}{2\pi}\,n\,s, \quad \varphi=\arccos \frac{\pi}{3n}\,,\quad  \gamma=\frac{\pi}{2n}\,.
\]
After some manipulations, we find:
\[
\sin\vartheta=\cos\varphi+\frac{2\pi}{3n}\cos\frac{\alpha}{2}\cos\gamma,\quad \cos\vartheta=\sin\varphi-\frac{2\pi}{3n}\cos\frac{\alpha}{2}\sin\gamma,
\]
whence
 \begin{equation}\label{titti}
 \tb_C=\cos\vartheta\cb_1-\sin\vartheta\cb_2,\quad  \tb_E=\cos\vartheta\cb_1+\sin\vartheta\cb_2,
 \end{equation}
with
\begin{equation}\label{teta}
\vartheta=\widehat\vartheta(n):=\arcsin\Big(\frac{\pi}{3n}\Big(1+\cos\frac{\pi}{2n}\Big)\Big);
\end{equation}
we note here for later use that $\,\sin(\vartheta-\gamma)>0$, for
all $n\geq 2$.

\vskip 6pt
\noindent{\textbf{(ii) \emph{Moment balance for the double ABU.}}}
The resultant moment of the force and couple system with respect to pole $P$ reads:
\begin{equation}\label{eqtors}
\mb_P+\bar\taub_{\alpha}(C)+\bar\taub_{\beta}(C)+\bar\taub_{\alpha
}(D)+\bar\taub_{\beta}(D)+\bar\taub_{\alpha}(E)+\bar\taub_{\beta
}(E)+\bar\taub_{\alpha
}(F)+\bar\taub_{\beta
}(F)=\mathbf{0}.
\end{equation}
Here, $\mb_P$ is the moment of the force system with respect to pole $P$:
$$
\mb_P=\pb_C\times\fb_C+\pb_D\times\fb_D+\pb_E\times\fb_E+\pb_F\times\fb_F,
$$
for which, with the use of $\eqref{posiz}$, \eqref{fortor}, and \eqref{altrepos}-\eqref{titti}, we find:
\begin{equation}\label{emmepi}
\mb_P=4\big(h\,s\sin\frac{\alpha}{2}\cos\vartheta+v\,\rho_0\sin\vartheta\big)\cb_2\,.
\end{equation}
Moreover,
\begin{equation}\label{manitau}
\begin{aligned}
&\bar\taub_{\alpha}(C)=\bar\tau_{\alpha}(C)\,(-\cb)\times\mb, \quad
\bar\tau_{\alpha}(C)=k_s\Delta\alpha(C)|\cb\times\mb|^{-1},\\
&\bar\taub_{\beta}(C)=-\bar\tau_{\beta}(C)\,\eb\times(-\cb),
\quad\bar\tau_{\beta}(C)=k_s\Delta\beta(C)|\eb\times\cb|^{-1},\\
&\bar\taub_{\alpha}(D)=-\bar\tau_{\alpha}(D)\,(-\cb)\times\mb, \quad
\bar\tau_{\alpha}(D)=k_s\Delta\alpha(D)|\cb\times\mb|^{-1},\\
&\bar\taub_{\beta}(D)=\bar\tau_{\beta}(D)\,(-\db)\times\eb,
\quad\bar\tau_{\beta}(D)=k_s\Delta\beta(D)|\db\times\eb|^{-1},\\
&\bar\taub_{\alpha}(E)=-\bar\tau_{\alpha}(E)\,\mb'\times(-\cb'), \quad
\bar\tau_{\alpha}(E)=k_s\Delta\alpha(E)|\mb'\times\cb'|^{-1},\\
&\bar\taub_{\beta}(E)=\bar\tau_{\beta}(E)\,(-\cb')\times(\eb'),
\quad\bar\tau_{\beta}(E)=k_s\Delta\beta(E)|\cb'\times\eb'|^{-1},\\
&\bar\taub_{\alpha}(F)=\bar\tau_{\alpha}(F)\,\mb'\times(-\cb'),
\quad
\bar\tau_{\alpha}(F)=k_s\Delta\alpha(F)|\mb^\prime\times\cb'|^{-1},\\
&\bar\taub_{\beta}(F)=-\bar\tau_{\beta}(F)\,(\eb')\times(\mb'),
\quad\bar\tau_{\beta}(F)=k_s\Delta\beta(F)|\eb'\times\mb'|^{-1},\\
\end{aligned}
\end{equation}
where
\begin{equation}\label{adsci}
\begin{aligned}
\eb&=-\cos 2\gamma\,\cb_1+\sin 2\gamma\,\cb_2,\quad \eb^\prime=\cos 2\gamma\,\cb_1+\sin 2\gamma\,\cb_2;\\
\cb^\prime&=\cos\frac{\alpha}{2}\ab^\prime+\sin\frac{\alpha}{2}\cb_3,\quad \db^\prime=\cos\frac{\alpha}{2}\ab^\prime-\sin\frac{\alpha}{2}\cb_3;\quad
\mb=-\db,\quad \mb^\prime=-\db^\prime.
\end{aligned}
\end{equation}
Making use of \eqref{emmepi}, \eqref{manitau} and \eqref{adsci}, it is not difficult to see that, as a consequence of the symmetries embodied in \eqref{coptorter},  \eqref{eqtors} is equivalent to the scalar equation \eqref{momentun}

\vskip 6pt
\noindent{\textbf{(iii) \emph{Moment balances for the $AC$ and $AD$ sticks.}}} For $A$ the common pole, we request that
\begin{equation}\label{eqtors1}
\begin{aligned}
\pb_C\times\fb_C+\bar\taub_\alpha(C)+\bar\taub_\beta(C)+\taub_\alpha(A)+\taub_\beta^u(A)=\mathbf{0},\\
\pb_D\times\fb_D+\bar\taub_\alpha(D)+\bar\taub_\beta(D)-\taub_\alpha(A)+\taub_\beta^d(A)=\mathbf{0},
\end{aligned}
\end{equation}
where the terms not listed in \eqref{manitau} are defined as follows:
\begin{equation}
\begin{aligned}
&\taub_\alpha(A)=\tau_\alpha(A)\,\cb\times\db, \quad
\tau_\alpha(A)=k_s\Delta\alpha(A)|\cb\times\db|^{-1},\\
&\taub_\beta^u(A)=-\tau_\beta^u(A)\bb\times\cb, \quad
\tau_\beta^u(A)=k_s\Delta\beta^u(A)|\bb\times\cb|^{-1},\\
&\taub_\beta^d(A)=\tau_\beta^d(A)\,\db\times\bb, \quad
\tau_\beta^d(A)=k_s\Delta\beta^d(A)|\bb\times\db|^{-1}.
\end{aligned}
\end{equation}
Then, equations \eqref{eqtors1} can be rewritten as
\begin{equation}\label{unadue}
\begin{aligned}
&\cb\times\Big(s\,\fb_C+\big(\bar\tau_\alpha(C)+\tau_\alpha(A)\big)\db-\bar\tau_\beta(C)\eb+\tau_\beta^u(A)\bb\Big)=\mathbf{0},\\
&\db\times\Big(s\,\fb_D+\big(\bar\tau_\alpha(D)-\tau_\alpha(A)\big)\cb-\bar\tau_\beta(D)\eb+\tau_\beta^d(A)\bb\Big)=\mathbf{0}.
\end{aligned}
\end{equation}
Note that $\Delta\alpha(A)$ is the fifth and last unknown in our equilibrium problem, the other four being $v$, $\Delta\alpha(C)$, $\Delta\beta(C)$, and $\Delta\beta^u$; the determining equations are \eqref{momentun} and the four scalar consequences of \eqref{unadue}: by subtraction of one of \eqref{unadue} from the other, we find both \eqref{delalf} and $\eqref{somma}_1$; by summation, we find $\eqref{somma}_2,_3$.

\subsection{Computations relative to Section \ref{rada}}\label{Arada}
Combining \eqref{eqpres} and \eqref{tauvarie}, we have:
\[
\cb\times\left(s\fb_C+2\tau_\alpha\db+\tau_\beta\bb-\bar\tau_\beta\eb\right)=\mathbf{0},
\]
an equation that, on recalling the definitions of the vectors involved, can be given the following form:
\[
s\fb_C+2\tau_\alpha\db+\tau_\beta\bb-\bar\tau_\beta\eb=\lambda\cb,\quad\textrm{with}\quad \lambda=-2\tau_\alpha,
\]
or rather, equivalently,
\begin{equation}
s\fb_C+2\tau_\alpha(\cb+\db)+\tau_\beta\bb-\bar\tau_\beta\eb=\mathbf{0}.
\end{equation}
The scalar consequences are two:
\[
\begin{aligned}
s(\fb_C\cdot\cb_1)-4\cos\frac{\alpha}{2}\cos\gamma\,\tau_\alpha+\tau_\beta+\cos 2\gamma\,\bar\tau_\beta=0,\\
s(\fb_C\cdot\cb_2)+4\cos\frac{\alpha}{2}\sin\gamma\,\tau_\alpha-\sin 2\gamma\,\bar\tau_\beta=0.
\end{aligned}
\]
If, in the second, we make use of \eqref{effeC} and \eqref{deltavari}, we find \eqref{prealfa}; with this, the first gives:
\begin{equation}\label{prefuno}
\fb_C\cdot\cb_1=-\frac{1}{2}\cot\gamma\sin\varphi\,p\,.
\end{equation}
 As to axial forces and strains, in the $a$-type stick $AB$ they
are, respectively,

\begin{equation}
-2\fb_C\cdot\bb=:N(AB)=\cot\gamma\sin\varphi p,
\end{equation}
whence we obtain \eqref{prea};
in all other $b$-type sticks they are:

\begin{equation}
\fb_C\cdot\cb:=N(AC)=\frac{1}{2}\Big(\cos\frac{\alpha}{2}\left(\sin\gamma+\cot\gamma\cos\gamma\right)\sin\varphi\Big)p,
\end{equation}
whence \eqref{preb} follows.

\subsection{Computations relative to Section \ref{325}}\label{A325}

Combining \eqref{eqpresz} and \eqref{tauvariez}, we have:
$$
\bb\times\big(s\fb_B+2\tau_\alpha\cb_3+\tau_\beta\db-\bar\tau_\beta\mb
\big)=\mathbf{0},
$$
an equation equivalent to
$$
s\fb_B+2\tau_\alpha(\cb_3+2\bb)+\tau_\beta(\db-\mb-2\bb)=\mathbf{0},
$$
whose two scalar consequences are:
\begin{equation}\label{scalz}
\left\{\begin{array}{c}s(\fb_B\cdot\cb_1)+4\tau_\alpha\sin\alpha-2\tau_\beta(\sin\alpha\cos\gamma+\sin\alpha)=0, \\\fb_B\cdot\cb_2=0.\qquad\qquad\qquad\qquad\qquad\qquad\qquad\qquad\end{array}\right.
\end{equation}
Equations \eqref{condz} and $\eqref{scalz}_2$ allow to conclude
that
$$
\fb_B\cdot\cb_1=\frac{1}{2\,{\sin\frac{\gamma}{2}}}\,p\,.
$$
With this, $\eqref{scalz}_1$, and \eqref{Zdiff}, we obtain \eqref{prealfaz}.

Finally, while it is obvious that $\Delta a=0$, for $\Delta b$ we have that
\begin{equation}
\Delta b=\frac{1}{k_a}\fb_B\cdot\bb=\frac{\sin\alpha}{2\sin\frac{\gamma}{2}}\frac{p}{k_a}.
\end{equation}

\subsection{Computations relative to Section \ref{62}}\label{4}
To arrive at relations \eqref{AsA} and \eqref{AnuA}, one starts from
\[
\begin{aligned}
&F=2n\,f\\
&H^{AM}=4\,b\sin\frac{\alpha}{2},\quad P^{AM}=2n\left(a+b\cos\frac{\alpha}{2}\right),\\
&\Delta H^{AM}=4\sin\frac{\alpha}{2}\Delta
b+2b\cos\frac{\alpha}{2}\Delta\alpha,\quad \Delta
P^{AM}=2n\left(\Delta a+\cos\frac{\alpha}{2}\Delta
b-\frac{1}{2}b\sin\frac{\alpha}{2}\Delta\alpha \right)
\end{aligned}
\]
(here, $a,b$ denote the common undeformed length of sticks of type $a$ and $b$, respectively), whence
\[
\begin{aligned}
\frac{\Delta H^{AM}}{H^{AM}}&=
\frac{1}{2\sqrt 3}\Big(3+\frac{1}{4}\left(1+\frac{3}{2\tan^2\beta} \right)^{-1}\frac{k_a s^2}{k_s} \Big)\frac{f}{k_a s},\\
\frac{\Delta P^{AM}}{P^{AM}}&= \frac{1}{2\sqrt
3}\Big(1-\frac{1}{4}\left(1+\frac{3}{2\tan^2\beta}
\right)^{-1}\frac{k_a s^2}{k_s} \Big) \frac{f}{k_a s}\,.
\end{aligned}
\]
Quite similarly, equations \eqref{ZsA} and \eqref{ZnuA} are arrived at on setting:
\[
\begin{aligned}
&F=n\,f\\
&H^{ZM}=2(a-b\cos\alpha),\quad P^{ZM}=2n\,b\sin\alpha,\\
&\Delta H^{ZM}=2(\Delta a-\cos\alpha\Delta
b+b\sin\alpha\Delta\alpha),\quad \Delta P^{ZM}=2n(\sin\alpha\Delta
b+b\cos\alpha\Delta\alpha),
\end{aligned}
\]
whence
\[
\begin{aligned}
\frac{\Delta H^{ZM}}{H^{ZM}}&=\left(\frac{3}{4}+\frac{3(1+\cos\beta)}{8(5+\cos\beta)}\frac{k_a s^2}{k_s}\right)\frac{f}{k_as}\,,\\
\frac{\Delta P^{ZM}}{P^{ZM}}&=
\left(\frac{1}{4}+\frac{\sqrt{3}(1+\cos\beta)}{8(5+\cos\beta)}
\frac{k_a s^2}{k_s}\right)\frac{f}{k_as}\,.
\end{aligned}
\]
\vskip 16pt
\centerline{*$\quad$*$\quad$*$\quad$*$\quad$*$\quad$*$\quad$*}
\vskip 16pt
It follows from \eqref{lamet} that
\[
w_{Rd}^A=2\,p^{-1} U^{ABU}_{Rd},
\]
where $U^{ABU}_{Rd}$ is deducible from \eqref{entor} by using
\eqref{prea} and \eqref{preb}:
$$
U^{ABU}_{Rd}=\frac{p^2}{16k_a}\,\frac{\sin^2\varphi}{\sin^2\gamma}\left(5+4\cos2\gamma+\frac{6}{8+3\left(\frac{\cos\gamma}{\sin\beta}\right)^2}
\,\frac{k_as^2}{k_s}\right).
$$
Thus,
$$
w_{Rd}^A=\frac{p}{8k_a}\,\frac{\sin^2\varphi}{\sin^2\gamma}\left(5+4\cos2\gamma+\frac{6}{8+3\left(\frac{\cos\gamma}{\sin\beta}\right)^2}
\,\frac{k_as^2}{k_s}\right).
$$

}

\end{document}